\documentclass[aip,jcp,preprint,longbibliography]{revtex4-1}
\usepackage{color}
\usepackage{amsmath}
\usepackage{amssymb}
\usepackage{graphicx}
\usepackage[T1, OT1]{fontenc}
\usepackage{hyperref}

\newcommand{\di}{\mathrm{d}}

\newcommand{\on}{\overline{n}}

\newcommand{\kon}{k_\mathrm{on}} 
\newcommand{\koff}{k_\mathrm{off}} 
\newcommand{\of}{{\overline f}}
\newcommand{\bnabl}{{\bf \nabla}}
\newcommand{\gd}{{\bf \nabla}}
\newcommand{\vf}{{\bf f}}

\begin{document}

\title{Bond formation kinetics affects self-assembly directed by ligand--receptor interactions\footnote{$^{\dag}$These authors contributed equally to this work; $^{*}$corresponding author: bmognett@ulb.ac.be}}

 \author{S.\ J.\ Bachmann$^\dag$, M.\ Petitzon$^\dag$, B.\ M.\ Mognetti$^*$}
 \affiliation{Universit\'e Libre de Bruxelles (ULB), Interdisciplinary Center for Nonlinear Phenomena and Complex Systems \& Service de Physique des Syst\`emes Complexes et M\'ecanique Statistique, Campus Plaine, CP 231,
Blvd du Triomphe, B-1050 Brussels, Belgium.}

\begin{abstract}
In this paper
we study aggregation kinetics in systems of particles functionalised by complementary linkers.
Most of the coarse--grained models currently employed to study { large--scale} self--assembly of these systems rely on effective
potentials between particles as calculated using equilibrium statistical mechanics.
In these approaches the kinetic aspects underlying 
the formation of  inter--particle 
% ligand--receptor 
linkages are neglected.
We show how the rate at which supramolecular linkages form drastically changes   
 the self--assembly pathway. 
In order to do this we develop a method that combines 
Brownian dynamics simulations 
% of the particle suspension 
with a Gillespie algorithm accounting for 
the evolution of inter--particle linkages.
If compared with dynamics based on effective potentials,
an explicit description of inter--particle linkages 
results in aggregates that in the early stages of self-assembly have 
 a lower valency.
Relaxation  towards equilibrium is hampered by the time required   
to break existing linkages within one cluster and {  to} reorient them toward free particles. 
This  effect is more important at low temperature and high particle diffusion constant. 
Our results highlight the importance of including kinetic rates 
into coarse-grained descriptions of ligand--receptor systems. 
\end{abstract}

\maketitle

\section{Introduction}
\begin{figure}
  \includegraphics[width=9cm]{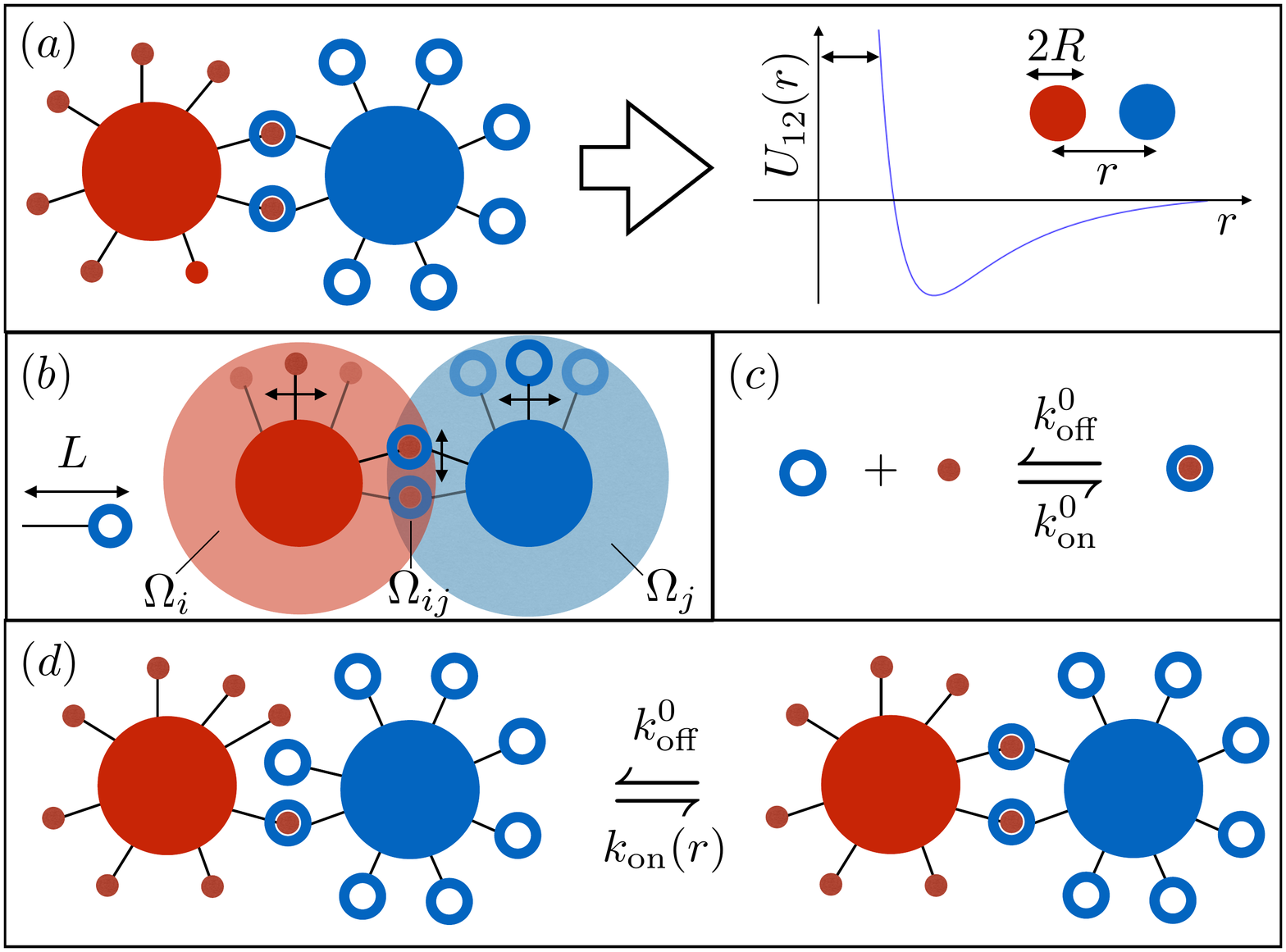} 
 \caption{(a) Particles functionalised with ligand-receptor constructs are usually modelled  by means of effective potentials $U_{12}$ that are calculated {  by} sampling over all possible inter-particle linkages. 
 (b) Considering ideal mobile linkers, $\Omega_i$ ($\Omega_j$) and $\Omega_{ij}$ are the volumes of the configurational space available to free and hybridised linkers.\cite{stefano-prl} The reaction rates of free linkers in solution (c) are modified in the 
 case of tethered constructs (d).  
  }
 \label{fig1}
\end{figure}

Supramolecular interactions mediated by ligand--receptor constructs 
are broadly used in nanoscience, for instance, to design systems that self--assemble\cite{kotera1994self,alivisatos,mirkin}
 or to develop materials for drug--delivery \cite{kiessling1} and remote--sensing \cite{taton} applications. 
Ligand-receptor interactions also control many biological functionalities including cell signalling and inter--membrane adhesion.
\\
{  Modelling} supramolecular interactions is challenging because it requires mapping atomistic details 
into coarse-grained representations of complex systems such as colloids\cite{hsu2010theoretical,mladek2012quantitative,mladek2,francrystals,lequieu2015molecular,knorowski2014self,dhakal2013growth,auyeung2014dna,stefano-prl,melting-theory2,melting-theory1,miriam,crocker-pnas,ding2014insights}   
or polymers\cite{DeGernier2014,dubacheva2015designing,xu2016simple} functionalised by reactive complexes. 
However, simplified models are necessary to access the time and length scales necessary to study, for instance, 
 selectivity in systems of functionalised {  nanoparticles},\cite{francisco-pnas} or self-assembly of large crystalline structures.\cite{dhakal2013growth,auyeung2014dna}
\\
A popular method of coarse-graining ligand-receptor systems relies on the use  
 of effective, state-dependent, potentials.
In this approach complex unit components, such as colloids functionalised by many polymeric strands,
are mapped into point--like particles
interacting {\em via} effective interactions (see Fig.\ \ref{fig1}a).
Effective interactions are calculated by sampling all possible micro--states involving 
two\cite{melting-theory2,melting-theory1,miriam,crocker-pnas,theodorakis} 
(or more\cite{stefano-prl}) particles at fixed {  distances}.\cite{mladek2012quantitative,mladek2,patrick-jcp,knorowski2014self,dhakal2013growth,auyeung2014dna}
Effective potentials have been largely used to study soft materials (e.g., in the blob representation of polymers 
\cite{bolhuis2001accurate}) and biological 
processes  (e.g., protein aggregation\cite{bieler2012connecting}).
\\
In this paper we question the use of effective potentials to study self-assembly of particles functionalised by 
ligands with mobile tethering point. 
We show that considering the finite rates at which supramolecular linkages form,
drastically changes self-assembly pathway. 
The effect is more important at low {  temperatures} where, already at the stage when only clusters with few particles are formed, 
most of the ligands are engaged in supramolecular linkages. 
The scarcity of free ligands hampers the formation of new bonds 
and therefore the growth of early--formed aggregates.  
Such reaction--limited dynamics not only slows down self-assembly but also results in  aggregates with low coordination number 
even at late stages of aggregations.
We classify different self-assembly kinetics 
using an experimentally accessible 
adimensional  parameter relating the time taken by particles to diffuse over a distance equal to the size of the ligands 
 with the average time required to form supramolecular linkages.
\\
The manuscript is organised as follows: In Sec.\ \ref{sec:model} we introduce the model and calculate the  rates  
of forming/breaking supramolecular linkages. 
In Sec.\ \ref{sec:simulation} we present and validate our simulation methods.
{ 
In Sec.\ \ref{sec:results} simulation results are reported while in Sec.\ \ref{sec:discussion} 
our findings are discussed.
}

%Coarse grained strategies that have been used to model self-assembly of functionalized particles rely on effective interactions in which all the degrees of freedom of the binders are integrated and their effect lumped into state-dependent interactions ($V$ in Fig.\ ???). 
%
%Effective interaction calculations are usually done at the pair of particles level but have also been used to study multi--body effects [Pair-Potential, Mladek, PRL].
%
%Effective interactions interactions could fail, for instance, at low binder density if the position of the binders, as in most cases, is not tracked. 
%

% Dynamics is also tricky and will be studied in the present contribution.

\section{The Model}\label{sec:model}

We study hard spheres of radius $R$ functionalized with $N$ linkers carrying reactive complexes that selectively bind (see Fig.\ \ref{fig1}).
In particular we consider a binary system made of two types of particles ($A$ and $\overline{A}$) 
functionalized by complementary linkers ($\alpha$ and $\overline{ \alpha}$, respectively). 
When two complexes hybridize an inter-particle linkage ({\em bridge}) forms. 
%\\
%{  To highlight the generality of our findings we also study single component systems featuring particles decorated by self-complementary linkers. For computational efficiency and clarity, the formation of intra-particle linkages ({\em loops}) is forbidden. }
%
\\
As found in systems of functionalized emulsions \cite{pontani2012biomimetic,feng2013specificity,hadorn2012specific} and lipid bilayers, \cite{mirjam-mobile,beales2014application,Banga2014,parolini2014thermal} as well as when modelling adhesion between biological membranes,  \cite{sackmann2014physics}
we consider linkers that can freely diffuse on  particle surfaces (Fig.\ \ref{fig1}b).   
In particular, we use the model presented in Ref.\ \cite{stefano-prl}, itself inspired by the experiments of van der Meulen 
and Leunissen in which solid colloids are coated with lipid bilayers.\cite{mirjam-mobile} 
Linkers are made of double--stranded DNA  anchored to the membrane by cholesterol anchors immersed into the bilayer, and
are tipped by reactive complexes like, for instance, short sequences of single--stranded DNA.  
In our modelling, such constructs are mapped into thin rigid rods of length $L$ terminated by reactive points that bind with 
a complementary partner under hybridization conditions (see Fig.\ \ref{fig1}a). 
We neglect rod-rod interactions but account for steric repulsions between hard particles and rods.\cite{stefano-prl}
In this work we consider $N_p=100$ particles of radius $R=5\cdot L$.\cite{stefano-prl} \\
{ 
In the following we denote particle spatial coordinates by $\{{\bf r}\}$ ($\{{\bf r}\}={\bf r}_1,\,{\bf r}_2,\,\cdots , {\bf r}_{N_p}$). 
The
}
hybridisation free energy of forming a given bridge between particles $i$ and $j$ at fixed $\{ {\bf r} \}$ 
is \cite{stefano-prl} 
\begin{eqnarray}
e^{-\beta \Delta G_{ij}(\{{\bf r}\})} =   \left( {\Omega_{ij}(\{{\bf r}\}) \over \rho_\ominus \Omega_i(\{{\bf r}\}) \Omega_j(\{{\bf r}\})} \right) e^{-\beta \Delta G_0}
\label{eq:DGhyb}
\end{eqnarray}
where $\Delta G_0$ is the hybridisation free energy of the reactive complexes when free in solution and $\rho_\ominus$ is the standard concentration ($\rho_\ominus=1\,$M). 
In Eq.\ \ref{eq:DGhyb}, $\Omega_{i}$ and $\Omega_{ij}$ are the configurational space volumes 
of a free linker on particle $i$ 
and of a bridge linking particle $i$ with particle $j$ (see Fig.\ \ref{fig1}b).  
Note that  $\Omega_{i}$ depends on the positions ${\bf r}_j$ of all particles $j$ that can be touched by linkers tethered on $i$ 
 while, in the low $L/R$ limit, $\Omega_{ij}$ is only a function of ${\bf r}_i$ and ${\bf r}_j$ (see Appendix \ref{sec:forcedecomposition}). 
\\
The linkage dynamics at given $\{ {\bf r} \}$ is specified by $\kon^{(i,j)}$ and  $\koff^{(i,j)}$: The rates at which bridges between particles $i$ and $j$ are formed and broken, respectively (see Fig.\ \ref{fig1}d). 
Using Eq.\ \ref{eq:DGhyb} and following Ref.\ \cite{parolini2016} it can be shown that 
%defining $\kon^0$ and $\koff^0$ the $on$ and $off$ rates of the sticky ends free in solution we have
%\begin{eqnarray}
%{\kon^0 \over \koff^0} = {\exp[-\beta \Delta G_0] \over \rho_\ominus}
%& \qquad &
%{\kon \over \koff} = \left( {\Omega_{ij}(\{{\bf r}\}) \over \rho_\ominus \Omega_i(\{{\bf r}\}) \Omega_j(\{{\bf r}\})} \right) {\exp[-\beta \Delta G_0] } \, .
%\label{eq:rates0}
%\end{eqnarray}
%
%Considering short DNA sticky ends a fair approximation allows to take $\koff=\koff^0$ [??] that when used in Eq.\ \ref{eq:rates0} becomes 
%
\begin{eqnarray}
& {\kon^{(i,j)} } = \left( {\Omega_{ij}(\{{\bf r}\}) \over \Omega_i(\{{\bf r}\}) \Omega_j(\{{\bf r}\})} \right) \kon^0 
 \qquad
\koff^{(i,j)} = {\rho_\ominus e^{\beta \Delta G_0} } \kon^0 , & 
\label{eq:rates}
\end{eqnarray}
where $\kon^0$ is the $on$ rate of reactive complexes when free in solution (see Fig.\ \ref{fig1}c), and we have assumed that the $off$ rates of free and 
tethered complexes are the same. \cite{parolini2016}  
Eqs.\ \ref{eq:rates} allow to calculate $\kon^{(i,j)}$ and $\koff^{(i,j)}$ for a given configuration $\{ {\bf r} \}$ as a function of 
$\kon^0$. 
Note that Eqs.\ \ref{eq:rates} neglect the fact that the mobility of linkers moving on particle surfaces is finite. \cite{mirjam-mobile}
This is a fair approximation in the case of DNA at low {  temperatures},
where the kinetics is dominated by DNA denaturation. \cite{parolini2016} 
% while their tethering point can diffuse for several micrometers in few seconds.\cite{mirjam-mobile}  
%
% In spite of the fact that for DNA sticky ends the $on$ rates are usually knows (for short oligomers  \cite{morrison1993sensitive,wetmur1968kinetics} $\kon^0 \approx 10^6\,\rho_\ominus^{-1}\,$s$^{-1}$),  in this work we will consider $\kon^0$ as a system parameter  and will highlight its key role in selecting self-assembly pathway. {\color{blue} actually kon w.r.t. diffusion } 
%
\\
Given the hybridisation free energies of single bond formation (Eq.\ \ref{eq:DGhyb}), we can derive the partition function of 
the system ($Z$).\cite{stefano-prl}
At finite particle density the configurational space available to free linkers is reduced by the presence of neighbouring colloids. 
This phase space contraction contributes to the partition function as (see Eq.\ \ref{eq:Z}) \cite{stefano-prl}
\begin{eqnarray}
Z_\mathrm{excl} (\{ {\bf r} \}) &=& 
\prod_{i=1}^{N_p}   
\left( {\Omega_i (\{ {\bf r}\}) \over \Omega_0} \right)^N \, ,
\end{eqnarray}
where $\Omega_0$ is the configurational space available to a free linker moving on an isolated colloid.
%Note that $Z_\mathrm{excl} $  is the partition function at fixed particle positions when $\Delta G_0 =\infty$ (see Eq.\ \ref{eq:DGhyb}).
Defining by $n_{ij} $ the number of bridges between particle $i$ and $j$ ($1\leq i<j\leq N_p$ and $\Delta G_0 <\infty$), the 
contribution to the partition function due to bridges at fixed $\{n_{ij} \}$  and $\{ {\bf r} \}$ is written 
{ 
as \cite{kitov2003nature,francisco-pnas,stefano-prl,xu2016simple} 
}
\begin{eqnarray}
Z_\mathrm{att} (\{ {\bf r} \}, \{ n_{ij} \}) &=& 
{ 
\prod_{i=1}^{N_p} {N! \over n_i ! \prod_q n_{iq}!} {  \cdot}\prod_{k<j} \left[n_{kj} ! e^{-\beta \Delta G_{kj}(\{ {\bf r} \}) n_{kj}}\right]
}
\nonumber \\
&=& \prod_{i=1}^{N_p} {N! \over n_i !} {  \cdot} \prod_{k<j} {e^{-\beta \Delta G_{kj}(\{ {\bf r} \}) n_{kj}}\over n_{kj} !}
\label{eq:Zatt}
\end{eqnarray}
where $n_i$ is the number of free linkers on particle $i$ ($n_i = N - \sum_j {n_{ij}}$). 
{ 
In the first line of Eq.\ \ref{eq:Zatt},  $N!/(n_i! \prod_q n_{iq}! )$ is the number of ways of 
{ 
partitioning $N$ linkers into $N_p$ groups $\{ n_{ip} \}$ ($p=1,\cdots N_p$) within which they are
indistinguishable,
}
 while $n_{kj}!$ is the number of ways of making $n_{kj}$ bridges starting from $n_{kj}$ complementary pairs.
}
%
% { 
% Note that $Z_\mathrm{att}=1$ (corresponding to a free energy contribution equal to zero) when bridges are not present ($n_{ij}=0$). 
% }
%
Finally the partition function of a $N_p$ particle system is obtained using $Z_\mathrm{excl}$ and $Z_\mathrm{att}$, 
defined above, and summing over all possible $\{n_{ij}\}$ and $\{ {\bf r} \}$ \cite{frenkel2014colloidal}
\begin{eqnarray}
Z &=&  {1\over {  (N_p/2)!^2} }  \int  \di \{ {\bf r}\} e^{-\beta V(\{ {\bf r} \}) }  Z_\mathrm{excl} (\{ {\bf r} \}) 
 \sum_{\{n_{  kj} \}} Z_\mathrm{att} (\{ {\bf r} \}, \{ n_{  kj} \}) 
\label{eq:Z}
\\
&=& \int { \di \{ {\bf r}\}  e^{-\beta V(\{ {\bf r} \}) } \over  {  (N_p/2)!^2}}  \sum_{\{n_{  kj} \}} \prod_{i=1}^{N_p}   
\left( {\Omega_i (\{ {\bf r}\}) \over \Omega_0} \right)^N {N ! \over n_i !}  \prod_{  k<j} {e^{-\beta \Delta G_{  kj}(\{ {\bf r} \}) n_{  kj}}\over n_{  kj} !} .
\nonumber
\end{eqnarray}
 In the latter expression the sum is taken over all possible {  bridges ($\{ n_{ij} \}$)} at given $\{{\bf r} \}$ (in particular $n_{ij}>0$ only if $i$ and $j$ are complementary and $0\leq n_i , n_{ij} \leq N$).
$V(\{{\bf r}\})$ is used to regularise hard-core interactions and  
has been taken equal to the sum of pairwise repulsions between two particles functionalized by 500 inert linkers of length $L_\mathrm{inert}=0.75 L$. This choice does not affect the conclusions of the paper and, if compared to sharper potentials, allowed to use larger integration steps. 
\\ 
% Different multivalent models (e.g.\ featuring loops or multilinker constructs [??]) can also be considered but they would require different combinatorial factors in Eq.\ \ref{eq:Z}.  
In Ref.\ \cite{stefano-prl} a saddle point approximation is used to replace the sum over $\{ n_{ij} \}$ in Eq.\ \ref{eq:Z} by their most 
probable values $\on_{ij}$. In particular in the large $N$ limit it can be shown that 
\begin{eqnarray}
Z = \int \di \{ {\bf r} \}   e^{ - \beta {\cal F} ( \on_{ij} , \beta \Delta G_{ij}(\{{ \bf r }\}), \Omega_i (\{ {\bf r} \})  ) } 
\label{eq:F}
\end{eqnarray}
with ${\cal F}$ defined by the integrand of Eq.\ \ref{eq:Z} and the most probable numbers of linkages $\on_{ij}$ satisfying chemical 
equilibrium equations \cite{stefano-prl}
\begin{eqnarray}
\on_{ij} = (N-\sum_{p} \on_{ip}) (N-\sum_{q} \on_{jq})  e^{-\beta \Delta G_{ij} (\{ {\bf r} \}) } \, .
\label{eq:saddle}
\end{eqnarray}  
If $\overline{n}_i = N - \sum_{p} \on_{ip}$ the previous equations can be rewritten as   
\begin{eqnarray}
\on_{i} &=& { N \over 1 + \sum_{p} \on_{p} \exp[-\beta \Delta G_{ip} ( \{ {\bf r} \} ) ] } 
\nonumber
\end{eqnarray}
that {  allow calculating} 
$\on_{i}$ and $\on_{ij}$ using self-consistent iteration methods.\cite{patrick-jcp}
%
% Remarkably when Eq.\ \ref{eq:saddle} is used to write $\{ \beta \Delta G_{ij} \}$ in terms of $\{\on_{pq}\}$ the free energy at the saddle point approximation assume the following remarkably compact form [??,??]
%
%\begin{eqnarray}
%\beta F &=& \beta  {\cal F} ( n_{ij} , \beta \Delta G_{ij}(\{{ \bf r }\},  \Omega_i (\{ {\bf r} \}) )|_{n_{ij}=\on_{ij}}
%\nonumber \\
%&=& \sum_i N \log \left( 1 - {1\over N } \sum_j \on_{ij} \right) +\sum_{i<j} \on_{ij} - N \sum_{i} \log {\Omega_{i} (\{ {\bf r} \}) \over \Omega_0 } \, .
%\label{eq:free}
%\end{eqnarray}
%
In Ref.\ \cite{stefano-prl} Eqs.\ \ref{eq:F} and \ref{eq:saddle}, together with portable expressions of ${\cal F}(\on)$, 
\cite{stefano-jcp,di2016communication} have been used to sample suspensions of functionalized particles by means of Monte Carlo 
algorithms in which the stationary numbers of linkages are 
calculated at each particle displacement using Eq.\ \ref{eq:saddle}. 
Note that because Eqs.\ \ref{eq:saddle} couple all $\{ \on_{ij} \}$ connecting particles that belong to the same cluster, ${\cal F}(\on)$ is multiboby.
The multibody nature of ${\cal F}$ is also due to the fact that  $\Omega_i(\{ {\bf r}\})$, controlling the formation of bridges 
between $i$ and $j$ (Eq.\ \ref{eq:DGhyb}), is function of all neighboring particles interacting with $i$ (see Appendix \ref{sec:forcedecomposition}). 
\\
In this work we study self-assembly kinetics by means of Brownian dynamics. 
We use an {\em implicit-linker method} in which forces between particles at given spatial configuration $\{ {\bf r} \}$ are evaluated 
by means of Eqs.\ \ref{eq:F} and \ref{eq:saddle}. In this scheme the number of linkers between particles are always taken 
equal to their most probable value (Eq.\ \ref{eq:saddle}) resulting in forces (${\bf \of}$) that are only function of $\{ {\bf r} \}$.
This is equivalent to having infinite $on$ rates ($\kon^0=\infty$).
We also develop an {\em explicit-linker method} in which we do not use Eq.\ \ref{eq:saddle} to instantaneously relax the numbers 
of bridges to their most probable values but treat $n_{ij}$ as dynamic variables. 
 In this case forces ${\bf f}$ are evaluated using the integrand of Eq.\ \ref{eq:Z} at given values of  
  $\{ n_{ij} \}$.
The number of bridges $\{ n_{ij} \}$ is updated using the Gillespie algorithm \cite{gillespie1977exact}
fed by $on$ and $off$ rates calculated using Eq.\ \ref{eq:rates}. 
\\
We first calculate the force acting on particle $i$ in the implicit-linker method (${\bf \of}_i$). 
Using Eqs.\ \ref{eq:Z}, \ref{eq:F} and \ref{eq:saddle}  
we obtain (defining ${\bf \nabla}_i = \partial/\partial {\bf r}_i$) \cite{patrickunpublished}
\begin{eqnarray}
 {\bf \of}_i &=& -\bnabl_i  {\cal F}( \on_{pq}  , \beta \Delta G_{pq}(\{ {\bf r} \}),  \Omega_p (\{ {\bf r} \})  )
\nonumber \\
&=& - \sum_{p<q} \bnabl_i \on_{pq} {\partial 
{\cal F}(  \on_{pq}  , \beta \Delta G_{pq}(\{{ \bf r} \}) ,  \Omega_p (\{ {\bf r} \})  ) 
\over \partial n_{pq} } -{\bf \nabla}_i  V(\{{\bf r} \})
\nonumber \\
 && - \sum_{p<q}  \bnabl_i   \Delta G_{pq} (\{ {\bf r} \}) {\partial   {\cal F}( \on_{pq}  , \beta \Delta G_{pq}(\{{ \bf r} \}) ,  \Omega_p (\{ {\bf r} \})  ) \over \partial \Delta G_{pq} (\{ {\bf r} \}) } 
 \nonumber \\
 && - \sum_{p} \bnabl_i  \Omega_{p} (\{ {\bf r} \}) {\partial   {\cal F}( \on_{pq}  , \beta \Delta G_{pq}(\{{ \bf r} \}) ,  \Omega_p (\{ {\bf r} \})  )
 \over \partial \Omega_{p} (\{ {\bf r} \}) } \, .
 \label{eq:calcf}
 \end{eqnarray}
 Noting that the saddle-point condition (Eq.\ \ref{eq:saddle}) 
 is written as $\partial {\cal F}(\on_{pq})/\partial n_{ij}=0$ and using the explicit form of ${\cal F}$ as 
 obtained comparing Eq. \ref{eq:Z} with Eq.\ \ref{eq:F} we find the following  expression  \cite{patrickunpublished}
\begin{eqnarray}
{\bf \of}_i = -\sum_{p<q} \on_{pq} \bnabl_i   \Delta G_{pq} (\{ {\bf r} \})  + N k_B T \sum_p { \bnabl_i  \Omega_p (\{ {\bf r} \}) \over  \Omega_p (\{ {\bf r} \})} -{\bf \nabla}_i  V(\{{\bf r} \}) \, .
\label{eq:force-implicit}
\end{eqnarray}
The expression of the forces in case of  explicit-linker dynamics is identical to Eq.\ \ref{eq:force-implicit}
when $\on_{ij}$ are replaced by $n_{ij}$. 
 In explicit--linker dynamics particle displacements are done at $n_{ij}$ constant and the first variation in the second line of Eq.\ \ref{eq:calcf} disappears resulting in
\begin{eqnarray}
{\bf f}_i = -\sum_{p<q} n_{pq} \bnabl_i   \Delta G_{pq} (\{ {\bf r} \})  + N k_B T \sum_p { \bnabl_i  \Omega_p (\{ {\bf r} \}) \over  \Omega_p (\{ {\bf r} \})} -{\bf \nabla}_i  V(\{{\bf r} \}) \, .
\label{eq:force-explicit}
\end{eqnarray}
Eqs.\ \ref{eq:force-implicit} and \ref{eq:force-explicit} will be used to develop Brownian dynamics simulations as described in Sec.\ \ref{sec:simulation}. 
A computationally efficient decomposition 
of Eqs.\ \ref{eq:force-implicit} and \ref{eq:force-explicit}, adapted to the system studied in this work, is reported in Appendix \ref{sec:forcedecomposition}.

 \begin{figure}
  \includegraphics[width=9cm]{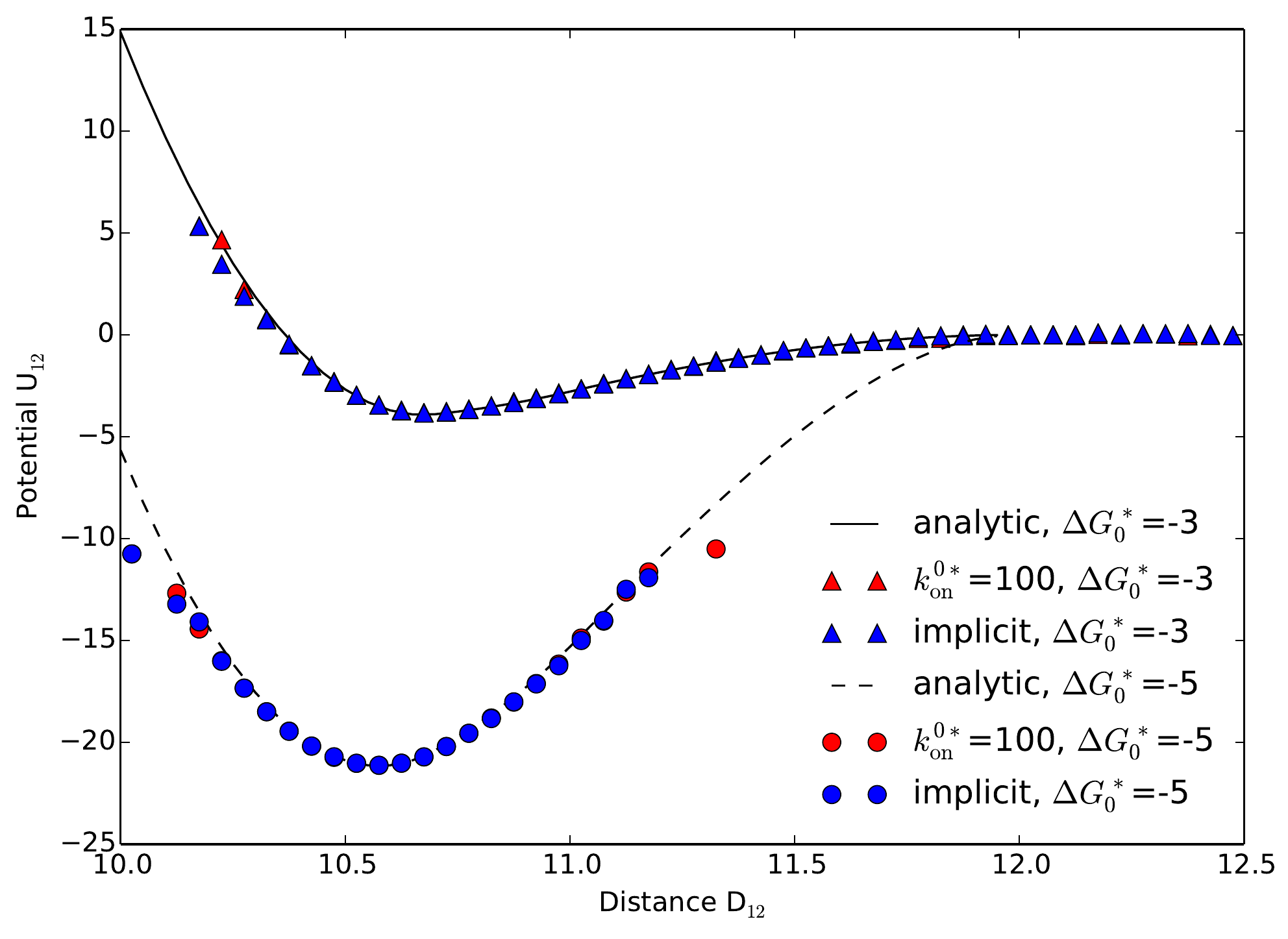} 
 \caption{
 Effective pair interactions as calculated using the  {\bf implicit--linker} (blue symbols) and the {\bf explicit--linker method} 
 (red symbols) as a function of the distance between centres of mass ($D_{12}$).
  Curves refer to analytic values of $U_{12}$ calculated using Eq.\ \ref{eq:F}.
 Two values of $\Delta G_0^*$ have been used. The simulated potentials have been shifted to match 
 analytic predictions at $2R^*+2$ when 
 $\Delta G_0^*=-3$, and at $r^*_\mathrm{min}$ (with $r^*_\mathrm{min}$ minimum of $U_{12}$) 
 when $\Delta G_0^*=-5$. 
 The same conclusions are reached when using different values of $\Delta t^*$. 
  }
 \label{fig:Veff}
\end{figure}
\section{Simulation methods}\label{sec:simulation}

Chosen simulation unit length and time are $L$ and $L^2/D$, where $D$ is the diffusion coefficient.
 In particular the adimensional $on$ rate of free complexes in solution is defined as  
$\kon^{0*}=\kon^0/(L D)$. Note that $\kon^{0*}$ can be interpreted as the ratio between the time taken by  
particles to diffuse a length equal to $L$ and the average time for a reaction between complexes in solution
at number density equal to $1/L^3$ to happen. 
In the following, all quantities tagged by $^*$ are meant to be in simulation units. For instance,  hybridisation free energies of tethered 
linkers (Eq.\ \ref{eq:DGhyb}) become 
\begin{eqnarray}
\beta \Delta G_{ij}(\{ {\bf r}^* \}) &=& \Delta G_0^* - \log \Big( {\Omega_{ij}^* ( \{ {\bf r}^* \} ) \over \Omega_{i}^* ( \{ {\bf r}^* \} )\Omega_{j}^* ( \{ {\bf r}^* \} )} \Big)
\nonumber \\
 \Delta G_0^* &=& \beta \Delta G_0 - \log{1\over \rho_\ominus L^3} \, \, .
\nonumber
\end{eqnarray}
\\
At each cycle of our algorithms we first update the number of bridges between particles at given $\{ {\bf r} \}$. 
In the { implicit-linker method} we numerically solve Eq.\ \ref{eq:saddle} and calculate $\on_{ij} (\{ {\bf r} \})$.
This is done by fixed point iterations as reported in Ref. \cite{patrick-jcp} and \cite{stefano-prl}. 
In the { explicit-linker method} we update $\{ n_{ij} \}$ using the Gillespie algorithm.\cite{gillespie1977exact}
We calculate the propensity of making/breaking a bridge between particles $i$ and $j$ as (see Eq.\ \ref{eq:rates})
\begin{eqnarray}
a_\mathrm{on}^{(i,j)} = n_i n_j \kon^{(i,j)} \qquad
a_\mathrm{off}^{(i,j)} = n_{ij} \koff^{(i,j)}
\label{eq:propensity}
\end{eqnarray}
and define 
$a_\mathrm{tot}=\sum_{i<j} a_\mathrm{on}^{(i,j)}+a_\mathrm{off}^{(i,j)}$. We sample the next reaction as well as the 
time interval for it to happen ($\tau$).
In particular we add (remove) a bridge between particles $i$ and $j$ with probability $a_\mathrm{on}^{(i,j)}/a_\mathrm{tot}$  
($a_\mathrm{off}^{(i,j)}/a_\mathrm{tot}$).
Note that at this stage we both sample the type of reaction and the particles involved.
$\tau$ is calculated by sampling from the probability distribution function 
$p(\tau)=a_\mathrm{tot} \exp[-a_\mathrm{tot} \tau]$.\cite{gillespie1977exact}
We then update the propensity matrices (Eqs.\ \ref{eq:propensity}) and repeat the entire process 
until the cumulative time interval of all reactions becomes bigger than the integration step $\Delta t^*$ 
used in the Brownian dynamics integration (see below). 
We note that more efficient algorithms, capable of updating the number of different ``species'' ($n_{ij}$) at time $t^*+\Delta t^*$ 
in one go, have been proposed.\cite{gillespie2007stochastic,rathinam2007reversible,grosfils2015role} 
Testing the efficiency of these schemes for the present system goes beyond the scope of this work.
\\
The second part of the algorithms is common to the two methods and consists in a Brownian dynamics step in which  
particles' centres of mass are evolved according to\cite{allen1989computer}
\begin{eqnarray}
{\bf r}^*_i (t+ \Delta t) = {\bf r}^*_i (t) + \beta {\bf F}_i L \Delta t^* + \sqrt{2 \Delta t^*} {\cal N}(0, 1 )
\label{eq:BD}
\end{eqnarray}
with ${\bf F}_i={\bf \of}_i$ (Eq.\ \ref{eq:force-implicit} and Appendix \ref{sec:forcedecomposition}) when using the 
{ implicit-linker method} and 
${\bf F}_i={\bf f}_i$ (Eq.\ \ref{eq:force-explicit} and Appendix \ref{sec:forcedecomposition}) when using 
the { explicit-linker method}. ${\cal N}(0,1)$ is a normally distributed random variable with zero mean and variance 
equal to one. In this work we use $\Delta t^*=0.001$.
{ 
Such value guaranteed affordable simulations and did not produce numerical instabilities usually found at high $\Delta t^*$. 
We also verified the fairness of the algorithm at small values of $\Delta t^*$ ($\Delta t^*=10^{-6}-10^{-4}$) 
by comparing effective pair interactions as done in Fig.\ \ref{fig:Veff}.
}
\\
The flow charts of the two algorithms are summarised below:\\ 
\begin{itemize}
\item{{\bf Implicit--linker method}}
\begin{itemize}
\item{while $t^*<t_\mathrm{tot}^*$}
\begin{itemize}
\item{for given  $\{ {\bf r}^*(t^*) \}$ calculate $\on_{ij}$ using Eq.\ \ref{eq:saddle} }
\item{calculate $\{ {\bf r}^*(t^*+\Delta t^*) \}$ using Eq.\ \ref{eq:BD} and ${\bf F}_i = {\bf \of}_i $ }
\item{$t^*\to t^*+\Delta t^*$}
\end{itemize}
\end{itemize}
\item{{\bf Explicit--linker method}}
\begin{itemize}
\item{while $t^*<t_\mathrm{tot}^*$}
\begin{itemize}
\item{for given  $\{ {\bf r}^*(t^*) \}$ calculate $a_\mathrm{off}^{(i,j)}$ and $a_\mathrm{on}^{(i,j)}$ using Eqs.\ \ref{eq:propensity} and Eqs.\ \ref{eq:rates}}
\item{while $\overline t^* < \Delta t^*$ }
\begin{itemize}
\item{sample the next reaction and the time interval $\tau$}
\item{upgrade $\{ n_{ij}\}$, $a_\mathrm{off}^{(i,j)}$, and $a_\mathrm{on}^{(i,j)}$ }
\item{$\overline t^* \to \overline t^*+\tau$ }
\end{itemize}
\item{calculate $\{ {\bf r}^*(t^*+\Delta t^*) \}$ using Eq.\ \ref{eq:BD} and ${\bf F}_i = {\bf f}_i $}
\item{$t^*\to t^*+\Delta t^*$}
\end{itemize}
\end{itemize}
\end{itemize}
{  In the $\Delta t^*\to 0$ limit,}
both methods satisfy detailed balance with respect to the distribution function $p(\{ {\bf r} \}) \sim \exp[-\beta {\cal F}(\{n_{ij}\}, \{ {\bf r} \} )]$ (Eq.\ \ref{eq:F}).\cite{rossky1978brownian}
In particular, they should reproduce the same results when used to sample equilibrium properties of the system.
This has been verified in Fig.\ \ref{fig:Veff} where, similar to what  done in experiments with particles in optical traps, \cite{crocker-pnas} we calculate effective pair interactions ($U_{12}$) by means of Boltzmann inversion.\cite{crocker-pnas} 
For two different values of $\Delta G_0^*$ and for $\kon^{0*}=100$ we verify that the two dynamics 
reproduce the same $U_{12}$ that also agree with the analytic results calculated using Eq.\ \ref{eq:F}. 
Note that in Fig.\ \ref{fig:Veff} we have chosen the lowest value of $\kon^{0*}$ used in the work.
This is the most subtle case in view of the fact that, at least for systems with two particles, the two dynamics become 
equivalent in the high $\kon^{0*}$ limit. 
\begin{figure*}[!ht]
%\vspace{-3.cm}
\begin{center}
\includegraphics[width=\textwidth]{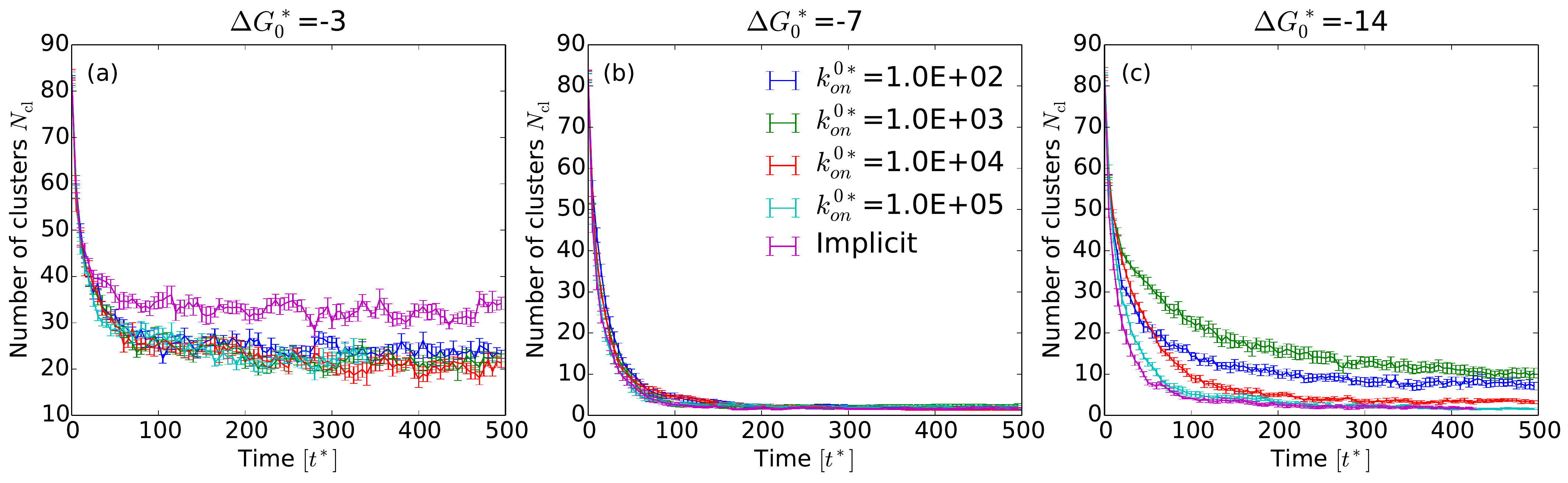} 
\end{center}
%\vspace{-2.8cm}
\caption{ 
Number of clusters {\em versus} simulation time at three different values of $\Delta G_0^*$.  
For { explicit--linker dynamics} at low values of $k_\mathrm{on}^{0*}$ and $\Delta G_0^*=-14$ particle aggregation is arrested. 
}\label{fig:cluster}
\end{figure*}
\section{Self-assembly kinetics}\label{sec:results}
\begin{figure}[h!]
%\vspace{-3.cm}
\begin{center}
\includegraphics[width=10.cm]{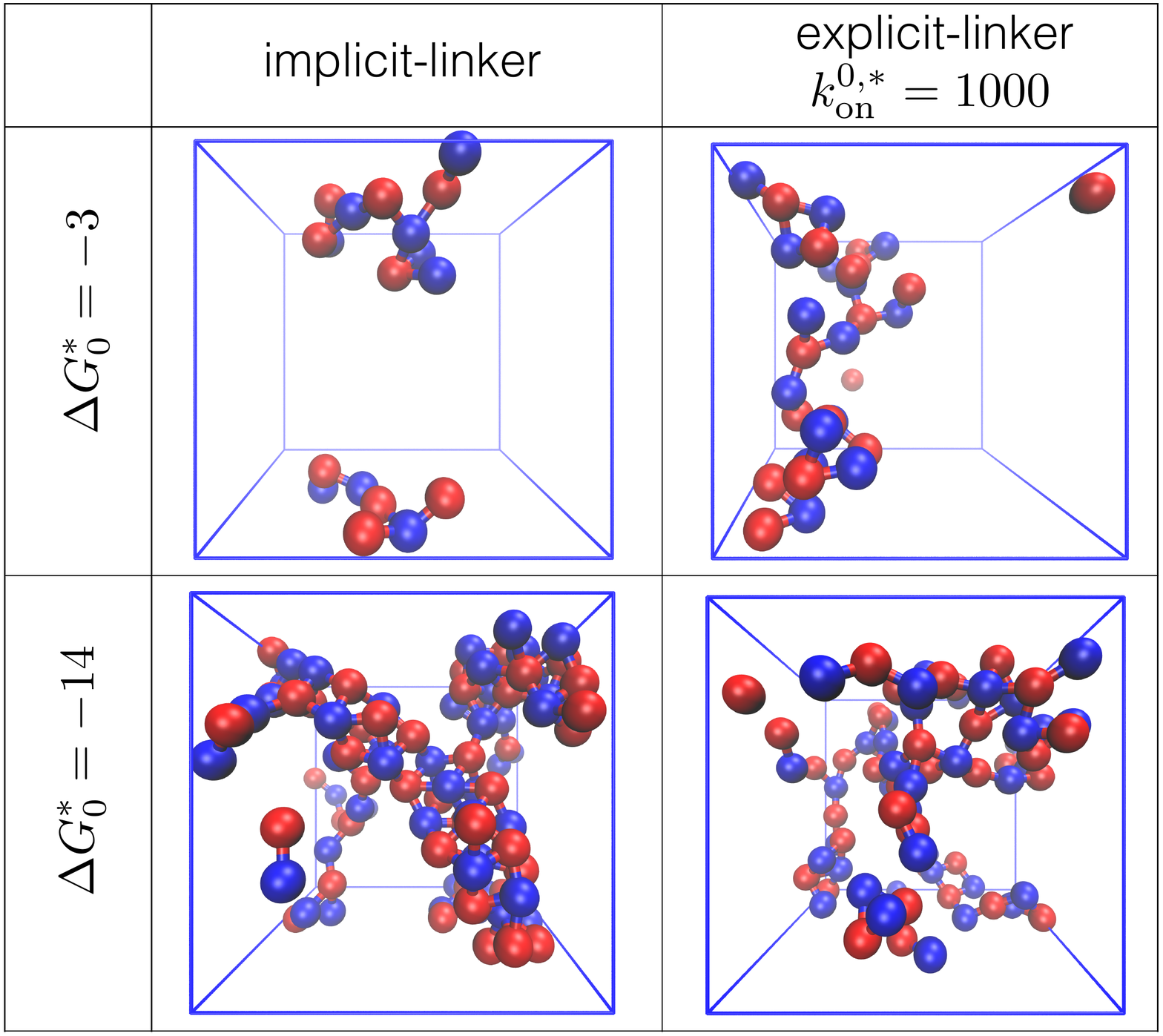} 
\end{center}
%\vspace{-2.8cm}
\caption{ 
Final simulation snapshots showing the biggest cluster.
For clarity particles in the gas phase, including smaller clusters, have not been reported.
For $\Delta G_0^*=-14$ the average coordination number is $\langle z \rangle \approx 2$ and  $\langle z \rangle  \approx 4$ for the explicit--linker and the implicit-linker method, respectively (see Fig.\ \ref{fig:neighbours}). 
}\label{fig:snapshots}
\end{figure}

\begin{figure}[h!]
%\vspace{-3.cm}
\begin{center}
\includegraphics[width=8.cm]{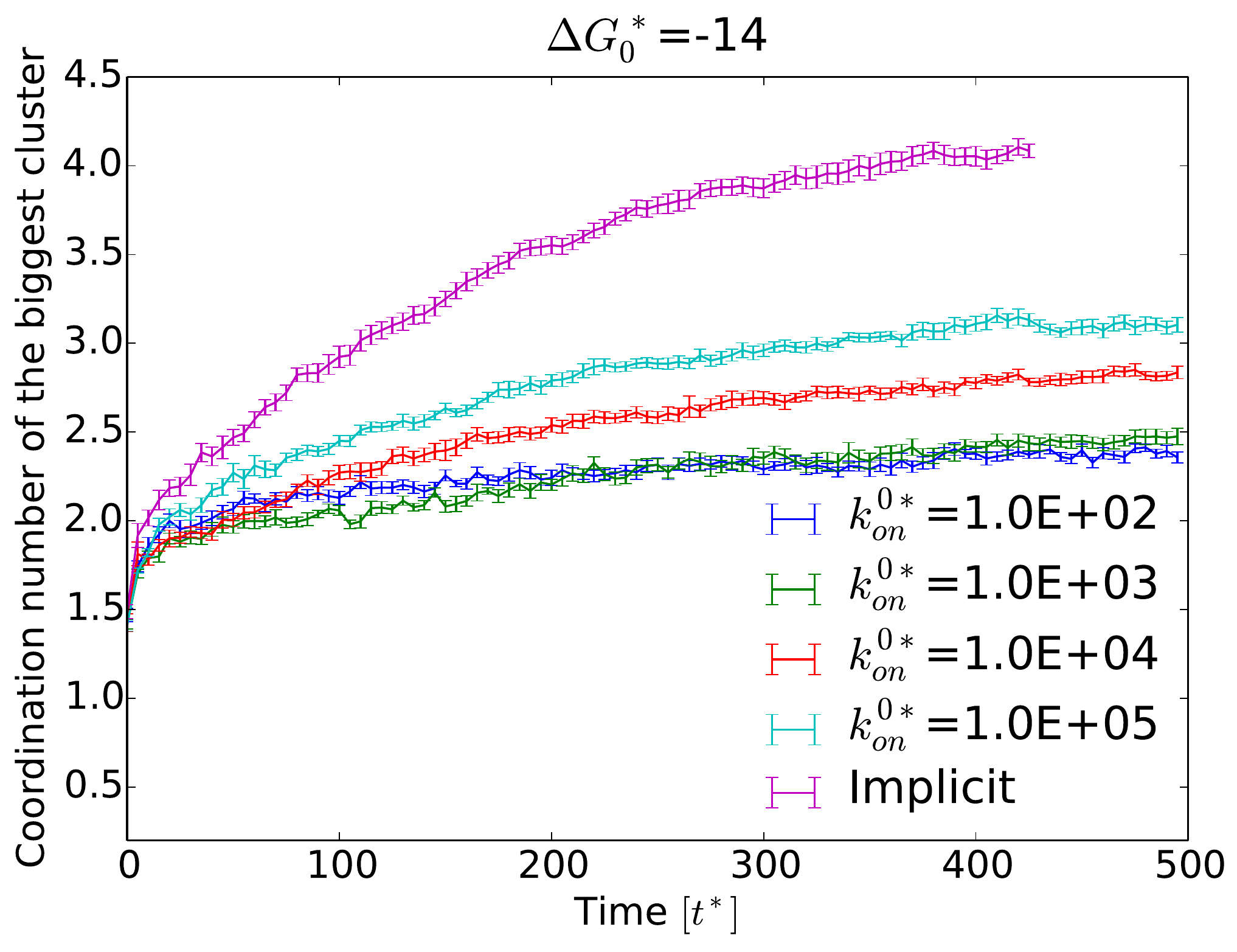} 
\end{center}
%\vspace{-2.8cm}
\caption{ 
Average valence ($\langle z \rangle$) of particles in the biggest cluster
 {\em versus} simulation time for $\Delta G_0^*=-14$. 
 %We compare trajectories obtained using 
%{\bf implicit--linker} and {\bf explicit--linker} simulations  with $k_\mathrm{on}^{0*}$ encompassing four orders of magnitude. 
%For {\bf explicit--linker dynamics} at low values of $k_\mathrm{on}^{0*}$ and $\Delta G_0^*=-14$ particle aggregation is arrested. 
%Each data set has been obtained by averaging ten independent trajectories. 
For explicit--linker simulations, valences tend to values slightly bigger than 
2 as confirmed by Fig.\ \ref{fig:neighbours}.
}\label{fig:neighbours}
\end{figure}

We simulate suspensions of $N_p=100$ colloids at packing fraction {  of} 0.1 starting from 
initial configurations in which particles are randomly distributed in the simulation box {  with periodic boundary 
conditions. }
We compare implicit--linker with explicit--linker simulations  with $k_\mathrm{on}^{0*}$ encompassing four orders of magnitudes.
Using {  the Stokes-Einstein} equation for the diffusion constants of spherical particles through water and $k_\mathrm{on}^{0}=10^6/$(M$\cdot$s), we 
estimate $k_\mathrm{on}^{0*}\approx   10^{-1} -   10^{0}$ for DNA coated colloids with radii ranging from 100$\,$nm to 1000$\,$nm.
{  This value is smaller than what considered in our study 
(see Figs.\ \ref{fig:cluster} and \ref{fig:neighbours}) because simulations at low values of $k_\mathrm{on}^{0*}$ are more expensive.
 Due to the fact that the kinetic bottlenecks described in this work are more severe at low $k_\mathrm{on}^{0*}$,
 our results highlight the importance of considering such effects in real experiments.
}
Moreover $k_\mathrm{on}^{0*}$ can be tuned across several orders of magnitude using other reactive complexes than DNA.
\cite{peck2015rapid}
In our simulations each data set has been obtained averaging over ten independent runs.
When using Python, each run required on average 7-14 days of an AMD Opteron 6274 CPU (2.2 GHz).
\\ 
In Fig.\ \ref{fig:cluster} we report the time evolution of the total number of clusters ($N_\mathrm{cl}$).
At $t^*=0$, $N_\mathrm{cl}$ is comparable with $N_p$. 
As aggregates start to form the number of clusters decreases until reaching a steady value. 
For $\Delta G_0^*=-3$, $N_\mathrm{cl}$ tends to $\approx 20-30$ (see Fig.\ \ref{fig:cluster}a). 
In this case a rich gas phase, consisting of around 10-20 singlets
(see SI Fig.\ S1a), few dimers and some trimers (see SI Fig.\ S2a,d), 
coexists with bigger clusters made of around 15-30 particles (see SI Fig.\ S3a).
These aggregates have a very low coordination number 
 ($\langle z \rangle \approx 1.9 - 2.5$, see SI Fig.\ S4a) and resemble branched colloidal chains (see Fig.\ \ref{fig:snapshots}).
Interestingly, 
{ 
in recent experiments of solid particles functionalised by mobile linkers
colloidal chains have been observed in a broad range of temperatures.\cite{mirjam-mobile}
}
\\
Surprisingly in Fig.\ \ref{fig:cluster}a results obtained with {  the two} different simulation methods do not perfectly agree.  
First we note that for $\Delta G_0^*=-3$ particles interact weakly (see Fig.\ \ref{fig:Veff})
resulting in reversible growth dynamics.
We also note that particle detachment rates 
are smaller when $\koff$ is finite than in implicit--linker dynamics.
Moreover, at fixed number of bridges, particles cross--linked by explicit linkers are closer than in the implicit linker case 
(note that the configurational free energy, Eq.\ \ref{eq:DGhyb}, diverges when the distance between particles is $2(R+L)$). 
It may be that, in explicit linker dynamics, particles can diffuse at the interface of aggregates
 long enough to stabilise themselves by binding 
a second colloid, most likely belonging to the same cluster.
%
% A particle cross-linked to more than one particle has less chances of detaching the cluster.
%
This mechanism may explain why aggregates grow faster when $\koff$ is finite.
 This hypothesis is supported by SI Fig.\ S4a showing 
that the average coordination number of aggregates is bigger in explicit--linker simulations. 
In SI Fig.\ S5 we repeated simulations using colloidal suspensions at lower density. 
In this case, smaller clusters are formed and the discrepancy between implicit and explicit 
simulations decreases. 
In SI Fig.\  S6 and S7 we test the precision of the saddle-point approximation by increasing the number of linkers to $N=100$.
First, in SI Fig.\  S6 we fine--tune the hybridisation free energy $\Delta G_0^*$ for the $N=100$ model 
resulting in a pair potential comparable with the $N=40$ model and $\Delta G_0^*=-3$. 
Using this model, in Fig. S7 we calculate the number of clusters versus time. 
{ 
The deviation between the two dynamics in Fig.\ S7 is comparable with what has been found in Fig.\ \ref{fig:cluster}a.
This shows that the saddle point approximation is not responsible for the discrepancies between implicit and explicit methods at 
 $\Delta G_0^*=-3$.
 A more detailed study of multibody effects in the reversible limit at different $\kon^{0*}$
  deserves further investigations.
}
\\
For $\Delta G_0^*\leq -7$ and for implicit--linker simulations particle attachment becomes irreversible and the gas phase 
tends to disappear.
In particular, for times bigger than $t^*\approx 100-150$ all particles are found in few clusters as expected 
deep in the coexisting region (see Figs.\ \ref{fig:cluster}b and \ref{fig:cluster}c).
At intermediate values of $\Delta G_0^*$ ($\Delta G_0^*=-7$ in Fig.\ \ref{fig:cluster}b)
the number of clusters calculated using  explicit--linker dynamics matches 
 implicit--linker results for any $\kon^{0*}$. 
 The same holds true for the number of singlets 
 (see SI Fig.\ S1b), the number of dimers and trimers (see SI Fig.\ S2b,e), the size of the biggest cluster (see SI Fig.\ S3b), and 
 for the average coordination number of the biggest cluster (see SI Fig.\ S4b).
\\
Intriguingly the two methods no longer agree at the lowest value of $\Delta G_0^*$ considered ($\Delta G_0^*=-14$ in Fig.\ \ref{fig:cluster}c)
where explicit--linker  simulations at low $\kon^{0*}$ {  fail} to assemble a single cluster.
In particular we find that for $\kon^{0*} <10^4$ significant amounts of dimers and trimers 
are still present at the end of {  the} simulations (see SI Fig.\ S2).
This implies that at low $\Delta G_0^*$
self-assembly is reaction--limited: The effective rate of attaching monomers to extended clusters is smaller than the same 
quantity at higher values of  $\Delta G_0^*$ (see discussion in Sec.\ \ref{sec:discussion}).
%
% The fact that the dynamics is reaction--limited is confirmed by the fact that for $\kon^{0*}\geq 10^5$ 
% the two methods predict similar assemblies. 
%
{
This is confirmed by SI Fig.\ S3c that reports the size of the biggest clusters. 
\\ 
{ 
The typical morphologies of the aggregates are also very different 
 when applying the two methods.
}
This is shown in 
Fig.\ \ref{fig:neighbours} where we study the average coordination number of the biggest cluster ($\langle z\rangle$) in the system
{\em versus} simulation time for $\Delta G_0^*=-14$. 
For explicit--linker dynamics, especially at low values of $\kon^{0,*}$, 
$\langle z \rangle$ increases very slowly  {  and remains} almost constant ($\langle z \rangle \approx 2$) 
during all simulations.
The typical biggest clusters for $\Delta G_0^*=-14$ are reported in the second {  row}  
of Fig.\ \ref{fig:snapshots}.  
For explicit--linker schemes we see that aggregates remain chain like even when they start 
to percolate through the simulation box. 
This proves that explicitly accounting for rate formation of supramolecular linkages not only slows down self--assembly 
but also results in more open aggregates. 
}

\begin{figure}[h!]
%\vspace{-3.cm}
\begin{center}
\includegraphics[width=8.cm]{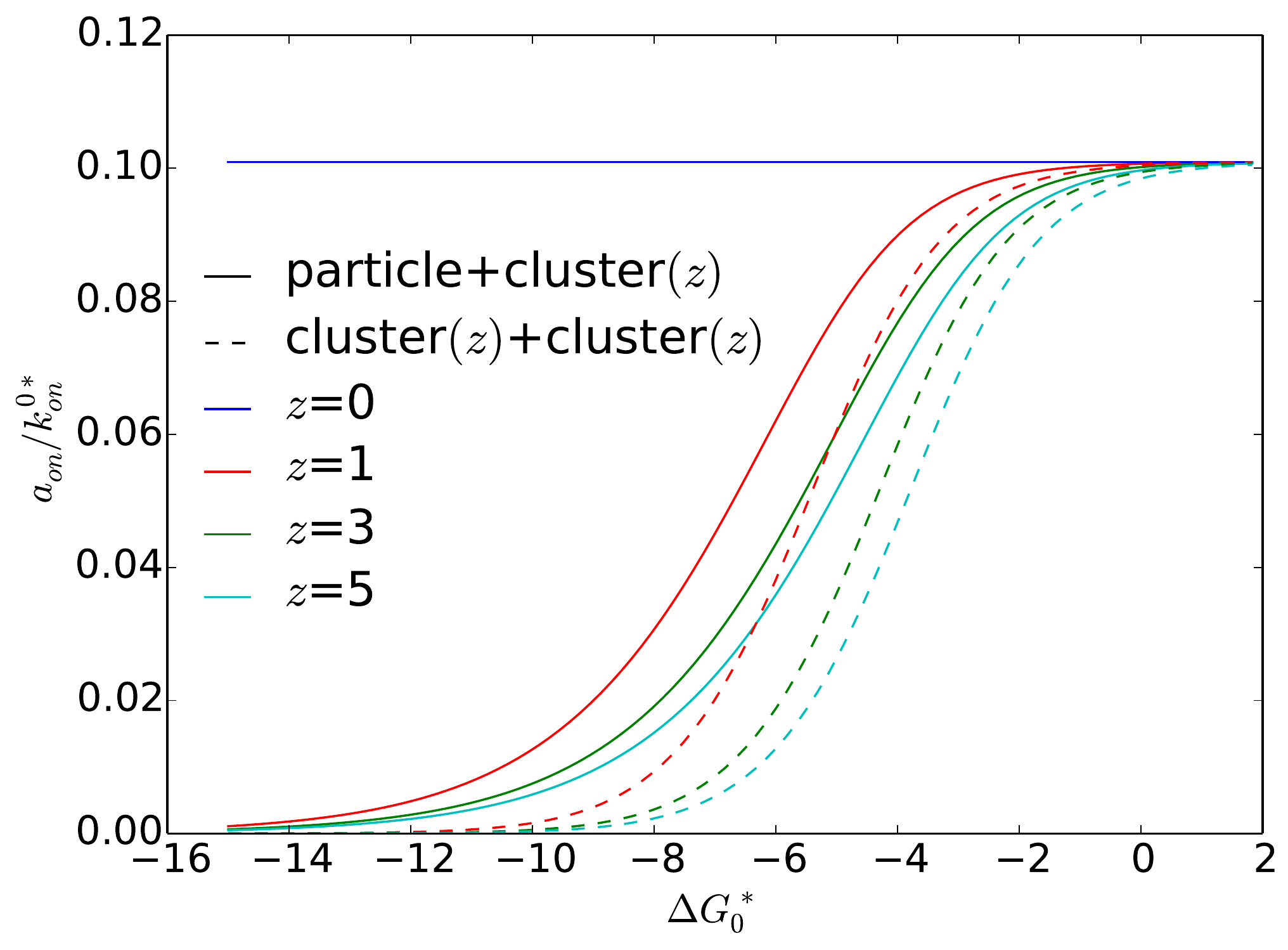} 
\end{center}
%\vspace{-2.8cm}
\caption{
Propensity (see Eq.\ \ref{eq:propensity}) of attaching an isolated particle to an existing cluster (solid lines), and propensity of creating a new linkage between two particles embedded in the same cluster (dashed lines).  In all calculations the distance between linkable particles has been taken equal to $2\cdot R+L$ and $z$ is the cluster coordination number.
}\label{fig:disc}
\end{figure}
\section{Discussion}\label{sec:discussion}

{ 
Self-assembly  of particles featuring explicit linkers 
is controlled by the finite time required  by existing clusters to capture a new particle.}
When using effective interactions, a particle wiggling around a cluster is immediately stabilized by 
the outer shell of the aggregate. 
On the other hand, {  in the cases} of models with explicit linker dynamics,
 available linkers on the clusters may not be fast enough to tie free particles letting them to diffuse away.
As shown in Fig.\ \ref{fig:cluster} this is more relevant at low $\Delta G_0^*$ because less free linkers are available 
and the rates of breaking  linkages becomes small (see Eq.\ \ref{eq:rates}).
This has been rationalized in Fig.\ \ref{fig:disc} (solid lines) where we calculate the propensity of attaching a singlet  
to a cluster as a function of $\Delta G_0^*$. 
In this figure 
we first consider infinite clusters with valence equal to $z$ and distance between bound particles equal to $2\cdot R+L$ 
and calculate the number of linkages per particle at equilibrium (Eq.\ \ref{eq:saddle}).
We then use Eq.\ \ref{eq:propensity} to calculate the propensity of making a bridge between one particle in the cluster and a free particle placed at distance equal to $2\cdot R+L$. While the propensity remains constant when linking two free particles ($z=0$), in case of extended clusters it drastically drops to zero for $\Delta G_0^*\lessapprox -10$. 
The effect is more important at high $z$ due to the fact that less free linkers are available in 
the presence of more neighbors.
This argument explains the results of Fig.\ {  \ref{fig:neighbours}}:
For $\Delta G_0^*=-14$ the propensity of attaching a singlet to an extended cluster 
{  become } negligible
and self-assembly is arrested. It also explains the fact that a significant amount of dimers is still present at 
the end of explicit--linker simulations (see SI Fig.\ S2c).
Plots similar to Fig.\ \ref{fig:disc} can be used to study the effects of changing the number of linkers $N$ or 
the size of the particles $R$ (see SI Fig.\ S8).
\\
Results similar to what is shown in Fig.\ \ref{fig:cluster} and Fig.\  \ref{fig:disc} have been reported in a study
of liposomes functionalized by two families of complementary DNA linkers.\cite{parolini2016}
In these systems inter--particle bridges compete with 
 intra--particle loops resulting in a slow--down of aggregation at low temperature.\cite{parolini2016} 
Linker sequestration may also underlay the findings of Ref.\ \cite{joshi2016kinetic} in which smart emulsions 
are used to adsorb controllable amounts of colloids without the need of reaching structural arrest. 
\\
Dashed lines in Fig.\ \ref{fig:disc} show the propensities of linking two particles, both of them belonging to a cluster with coordination 
number equal to $z$ (see Eq.\ \ref{eq:propensity}).
Also in this case $a_\mathrm{on}$ goes to zero at low $\Delta G_0^*$. 
However now the effect is more important in view of the fact that both particles feature linkers that are already engaged
in pre-existing linkages.
%
%This explains the results of Figs.\ {\color{blue} ?? ??} where, for intermediate $\Delta G_0^*$, 
%we observe the same number of clusters (Fig.\ {\color{blue} ??}) but with an average number of neighbours that 
%is lower than in the implicit--linker method (Fig.\ {\color{blue} ??}). 
%
{
These results explain the fact that in explicit--linker simulations chain--like structures do not fold onto 
themselves but persist until
they percolate through the simulation box.  
}
\\
In conclusion we have studied self--assembly of particles functionalized by mobile ligands
and have proven how accounting for linkage dynamics slows down colloid aggregation and results 
in clusters with low valence if compared with dynamics based on effective potentials. 
{ We believe our findings will be useful to develop coarse--grained models 
of supramolecular systems.
}

{\bf Acknowledgements} SJB and BMM are supported by the Universit\'e Libre de Bruxelles (ULB). 
We acknowledge S.\ Angioletti-Uberti, L.\ Di Michele, {  P.\ Gaspard,} and S.\ Napolitano for discussions. 
We acknowledge an anonymous reviewer 
for suggesting a more direct derivation of Eq.\ \ref{eq:appforcefinal} than what was presented in an early version of the manuscript.
Computational resources have been provided by the Consortium des {\'{E}}quipements de Calcul Intensif (C{\'{E}}CI), funded by the Fonds de la Recherche Scientifique de Belgique (F.R.S.-FNRS) under Grant No. 2.5020.11.
%{\color{blue} Add the link to the code somewhere.}

\appendix

\section{Explicit calculation of ${\bf f}_i$ and ${\bf \of}_i$ }\label{sec:forcedecomposition}

\begin{figure}
\vspace{0.cm}
\includegraphics[width=8.cm]{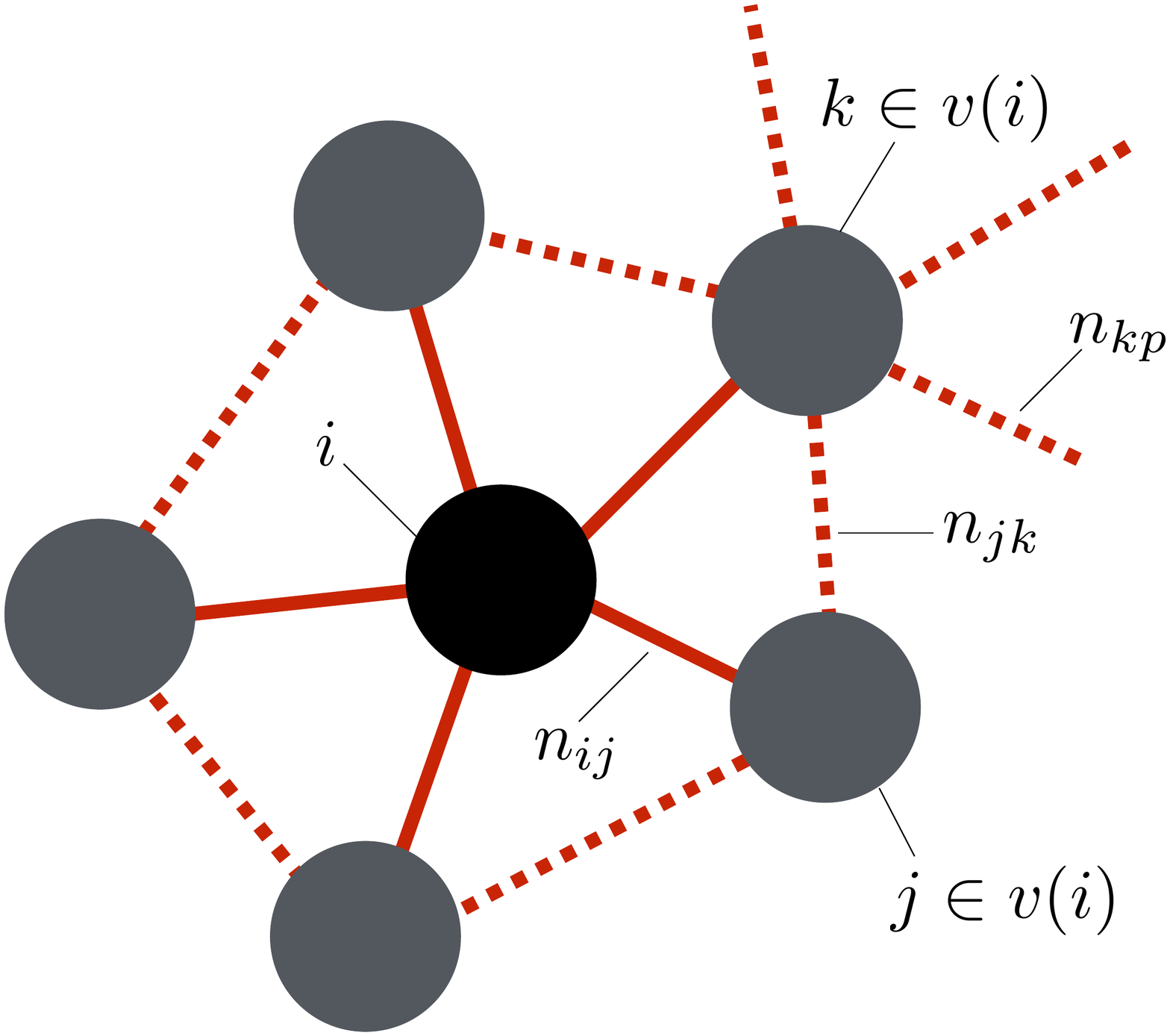} 
\caption{
The force acting on a given particle $i$ is only due to particle $j$ ($j\in v(i)$) that can directly interact with $i$
(see Eq.\ \ref{eq:appforcefinal}). 
}\label{fig:app}
 \vspace{-0.cm}
 \end{figure}

In this appendix we further develop Eqs.\ \ref{eq:force-explicit} and \ref{eq:force-implicit} and derive tractable expressions that can 
be used in the Brownian dynamics algorithms described in Sec.\ \ref{sec:simulation}.
Using Eq.\ \ref{eq:DGhyb} and Eq.\ \ref{eq:force-explicit} we obtain 
\begin{eqnarray}
 \beta {\bf f}_i  &=& \sum_{p<q} n_{pq} \bnabl_i \log {\Omega_{pq}(\{ {\bf r} \}) \over \rho_\ominus \Omega_{p}(\{ {\bf r} \})\Omega_{q}(\{ {\bf r} \})}    + N \sum_p { \bnabl_i  \Omega_p (\{ {\bf r} \}) \over  \Omega_p (\{ {\bf r} \})}  
\nonumber \\
&& - \beta {\bf \nabla}_i  V(\{{\bf r} \}) \, \,  .
\label{eq:forceapp1a}
\end{eqnarray}
Using $N=n_i + \sum_j n_{ij}$ the previous relation can be further simplified as follows
\begin{eqnarray}
 \beta {\bf f}_i  &=& \sum_{p<q} n_{pq}  { \bnabl_i \Omega_{pq}(\{ {\bf r} \}) \over \Omega_{pq}(\{ {\bf r} \}) }  
   +  \sum_p n_p { \bnabl_i  \Omega_p (\{ {\bf r} \}) \over  \Omega_p (\{ {\bf r} \})}   - \beta {\bf \nabla}_i  V(\{{\bf r} \}) \, \,  .
\nonumber \\
\label{eq:forceapp1}
\end{eqnarray}
Eq.\ \ref{eq:forceapp1} is computationally more tractable than Eq.\ \ref{eq:forceapp1a} because it allows calculating 
the force on particle $i$ by cycling {  over} all particles $j$ that interact with $i$, either by forming bridges or by linker-colloid repulsion 
(see particles jointed to $i$ by full curves in Fig.\ \ref{fig:app}).
Instead, attempting to evaluate the first term of Eq.\ \ref{eq:forceapp1a} would also require to cycle on second-neighbour particles
(see dashed lines in Fig.\ \ref{fig:app}).
\\
In the limit in which the length of the {  mobile} linkers is much smaller than the radius of the colloids 
($L/R \ll 1$) 
the reactive end-points of the 
linkers (see Fig.\ \ref{fig1}) are uniformly distributed within the shell of radii $R$ and $R+L$.
In such limit $\Omega_i$ and $\Omega_{ij}$ (below denoted as $w_{ij}$) are equal to the volume available to free and hybridized end-points, respectively 
(the {  reliability} of this approximation has been tested in the Supplementary Informations of Ref.\ \cite{stefano-prl}).
If $e_{ij}$ is the volume excluded to a free linker on particle $i$ due to the presence of particle $j$, in the limit $L/R \ll 1$ we have
\begin{eqnarray}
\Omega_i = \Omega_0 - \sum_{j\in v(i)} e_{ij}
& \qquad &
\Omega_{ij} = w_{ij} 
\label{eq:forceapp2}
\end{eqnarray}
where
 $v(i)$ is the ensemble of particles 
 which linkers can touch particle $i$ (see Fig.\ \ref{fig:app}) and $\Omega_0=4 \pi R^2 L$. 
 {  In Eq.\ \ref{eq:forceapp2} }
 $w_{ij}$ and $e_{ij}$ are given by the following {  expressions} 
\begin{eqnarray}
e_{ij}(r) &=& V_\mathrm{ovl} (R+L,R,r)
\\
w_{ij}(r) &=& V_\mathrm{ovl} (R+L,R+L,r) - 2 V_\mathrm{ovl} (R+L,R,r) \, ,
\label{eq:w}
\end{eqnarray}
where $r=|{\bf r}_i -{\bf r}_j|$ and $V_\mathrm{ovl} (R_1,R_2,r) $ is the overlapping volume of two spheres of radius $R_1$
and $R_2$ placed at center-to-center distance equal to $r$ 
\begin{eqnarray}
V_\mathrm{ovl} (R_1,R_2,r) &=& {\pi \over 12 r} (D-r)^2(r^2+2r D -3d^2) \, \, ,
\label{eq:vovl}
\end{eqnarray}
with $d\leq r\leq D$, $d=|R_2-R_1|$, and $D=R_1+R_2$.
{ 
Eqs.\ \ref{eq:w} and \ref{eq:vovl} imply that bridges can interact only with the two particles to which they are tethered. 
}
{  Inserting  Eqs.\ }\ref{eq:forceapp2} {  into} Eq.\ \ref{eq:forceapp1} we obtain our final result
\begin{eqnarray}
\beta \vf_i  &=& \sum_{j} n_{ij}  {\gd_i w_{ij} \over w_{ij} } -  \sum_{j\in v(i) } \left[ n_i   {\gd_i  e_{ij} \over \Omega_i } +  n_j  {\gd_i e_{ji} \over \Omega_j } \right]  - \beta {\bf \nabla}_i  V(\{{\bf r} \}) \, .
\nonumber \\
\label{eq:appforcefinal}
\end{eqnarray}
Using Eq.\ \ref{eq:appforcefinal} we calculate the force acting on a given particle $i$ using a single loop over all 
{ 
particles belonging to $v(i)$
}
 and tracking the values of $\Omega_j$  $\forall\,\, j$.
The expression of ${\bf \of}_i$ is obtained 
replacing $n_{ij}$ and $n_p$ with $\on_{ij}$ and $\on_p$ in Eq.\ \ref{eq:appforcefinal}.

\section{Supplementary Material: Bond formation kinetics affects self-assembly directed by ligand--receptor interactions}

%\maketitle

 \begin{figure}[h]
    \includegraphics[width=17cm]{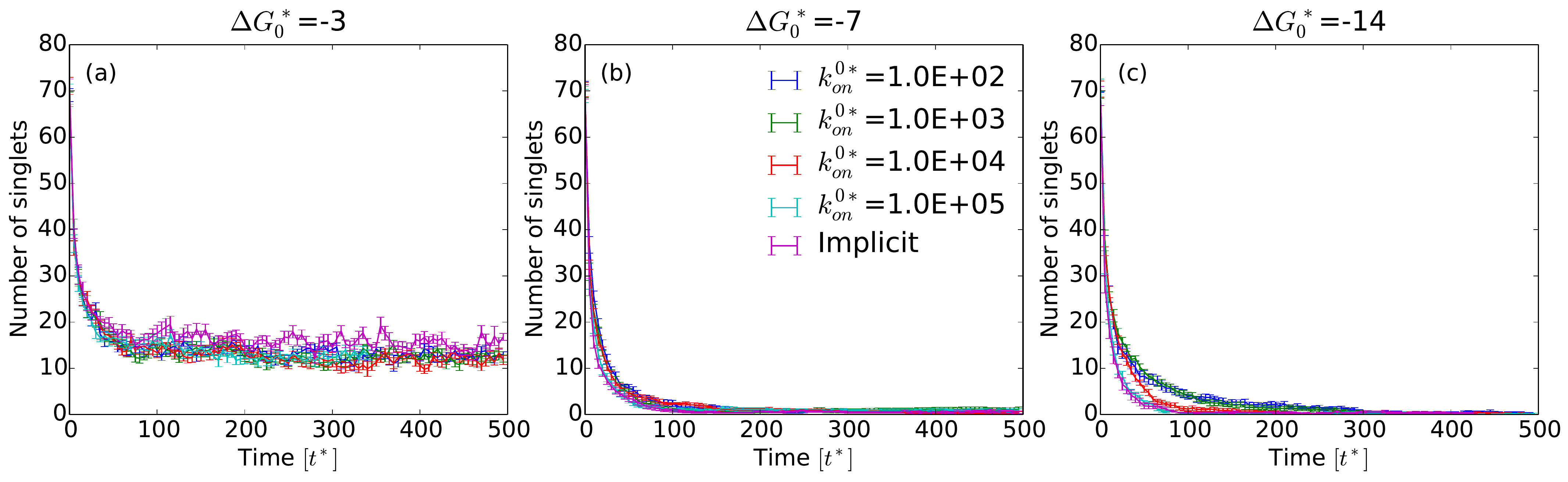} 
 \caption{Number of free particles \textit{versus} simulation time at three different values of $\Delta G_{0}^{*}$. 
 %for the implicit simulation (purple) and for four different explicit simulations with different $k_{on}^{0*}$.
 }
 \label{fig:S0}
\end{figure}

  \begin{figure}[h]
    \includegraphics[width=17cm]{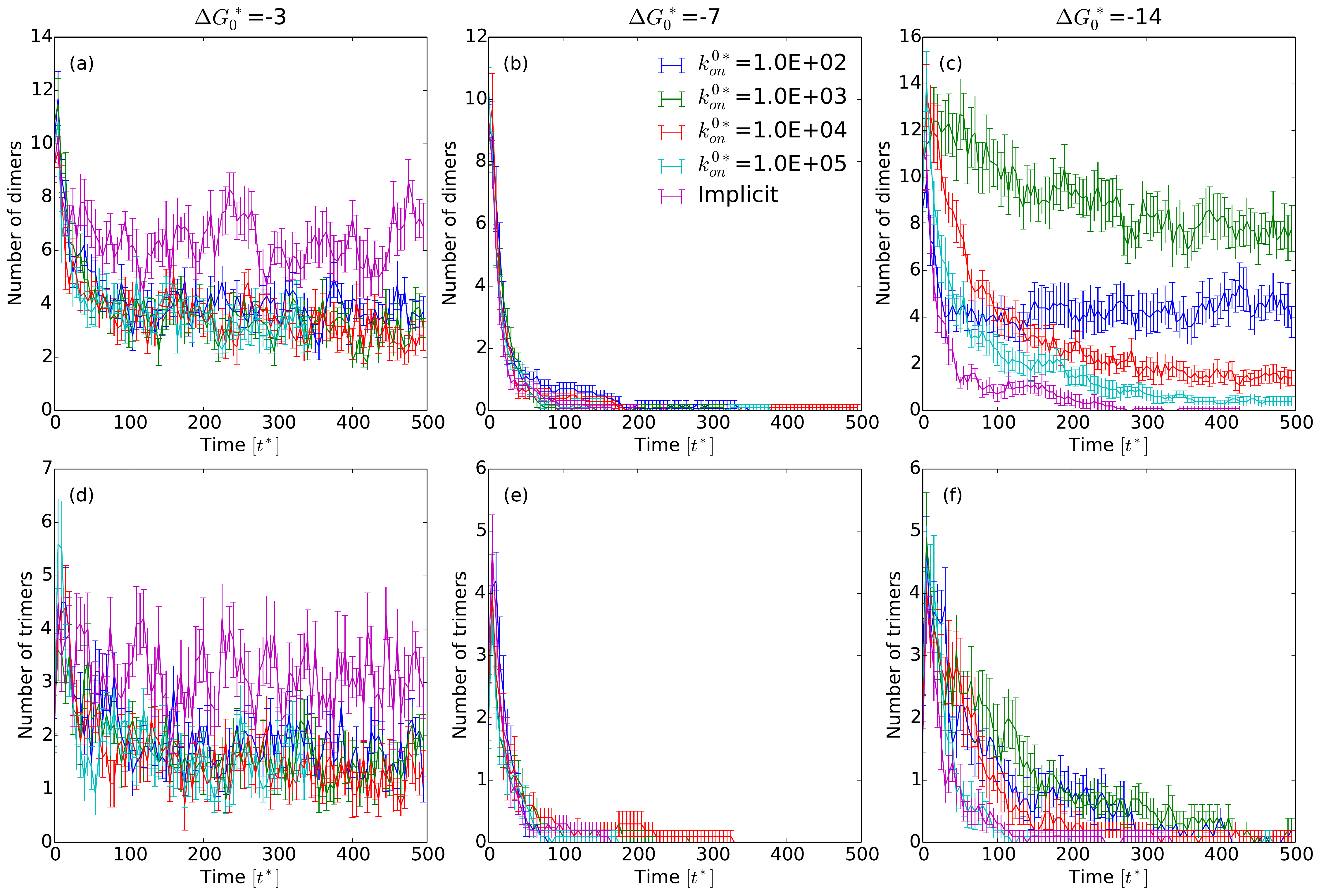}
 \caption{Number of dimers (first row) and trimers (second row) \textit{versus} simulation time at three different values of $\Delta G_{0}^{*}$. %for the implicit simulation (purple) and for four different explicit simulations with different $k_{on}^{0*}$.
 }
 \label{fig:S1}
\end{figure}

  \begin{figure}[h]
    \includegraphics[width=17cm]{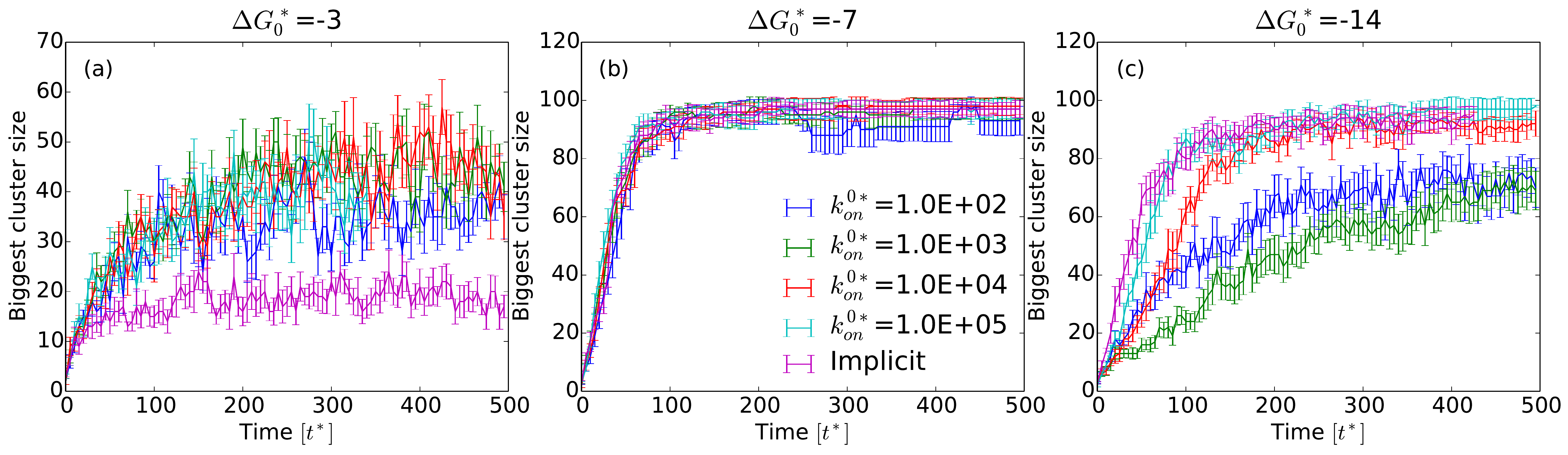} 
 \caption{Size of the biggest cluster in the system \textit{versus} simulation time at three different values of $\Delta G_{0}^{*}$.
 % for the implicit simulation (purple) and for four different explicit simulations with different $k_{on}^{0*}$.
 }
 \label{fig:S2}
 \end{figure}

 \begin{figure}[h]
    \includegraphics[width=17cm]{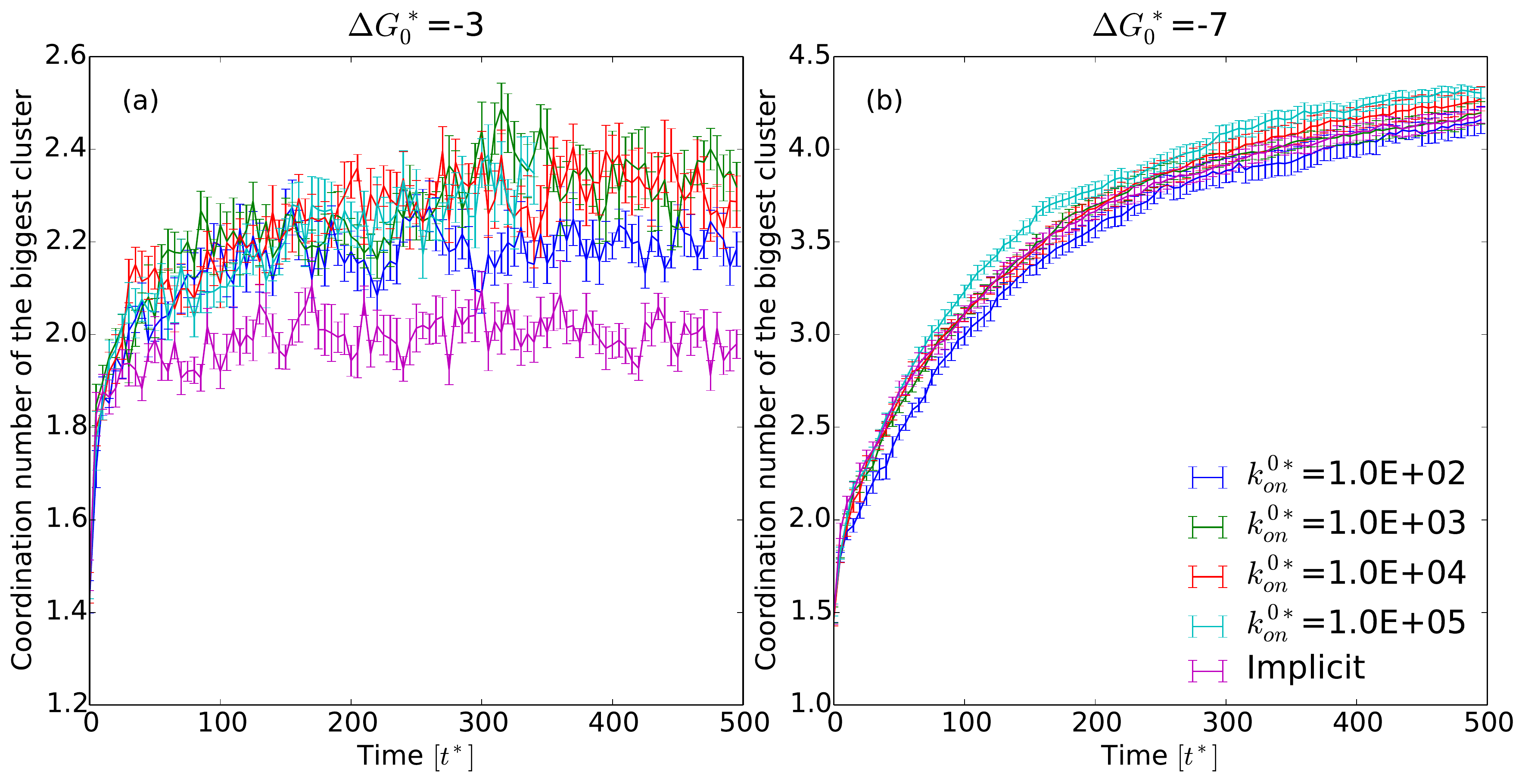} 
 \caption{Average valency $\left(\left\langle z \right\rangle \right)$ of particles in the biggest cluster \textit{versus} simulation time for two different $\Delta G_{0}^{*}$.
 % and for the implicit simulation (purple) and for four different explicit simulations with different $k_{on}^{0*}$.
 }
 \label{fig:S3}
\end{figure}

 \begin{figure}[h]
    \includegraphics[width=17cm]{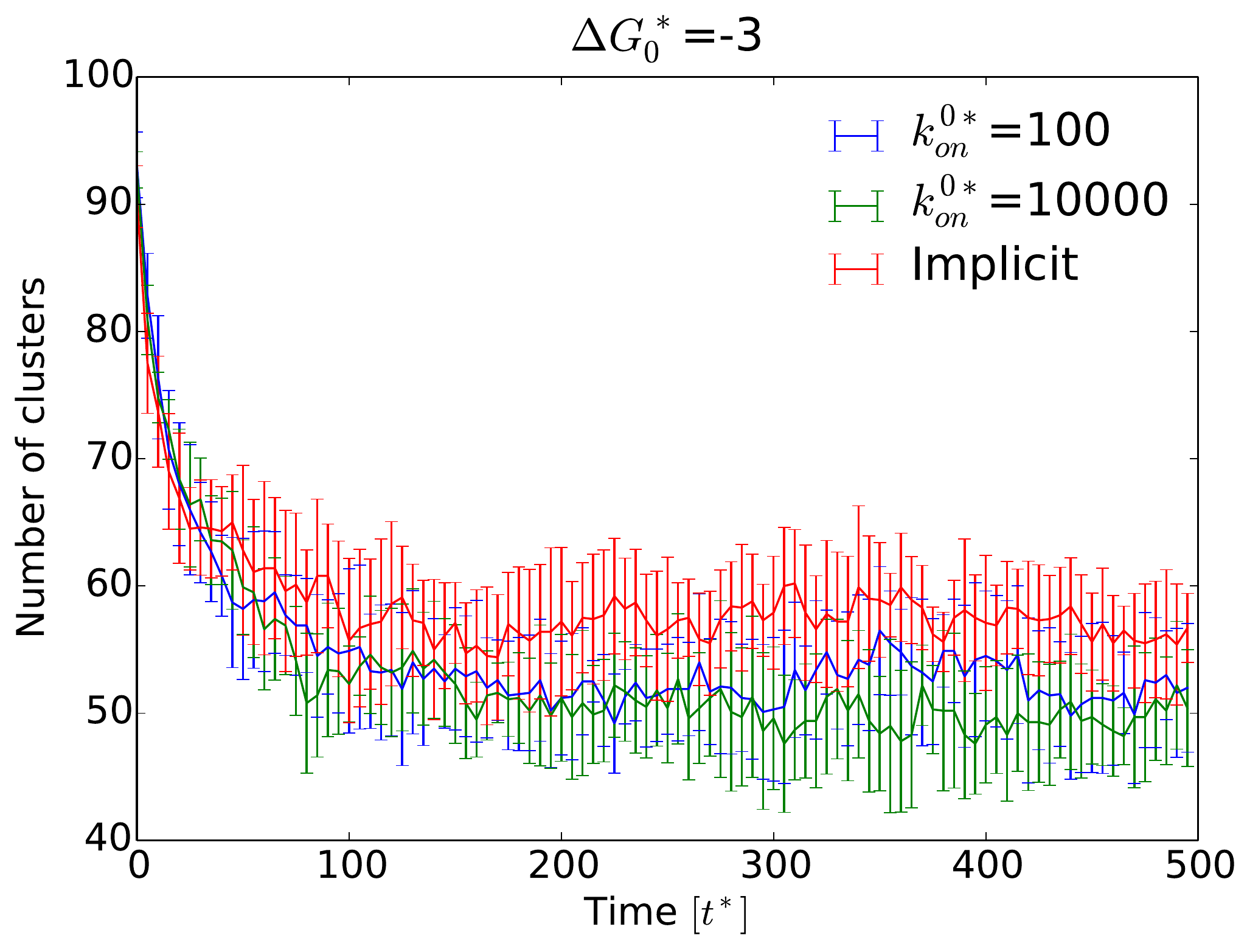} 
 \caption{Number of clusters {\em versus} simulation time at packing fraction 0.05 and $\Delta G_{0}^{*}=-3$.
 }
 \label{fig:S4}
\end{figure}

 \begin{figure}[h]
    \includegraphics[width=17cm]{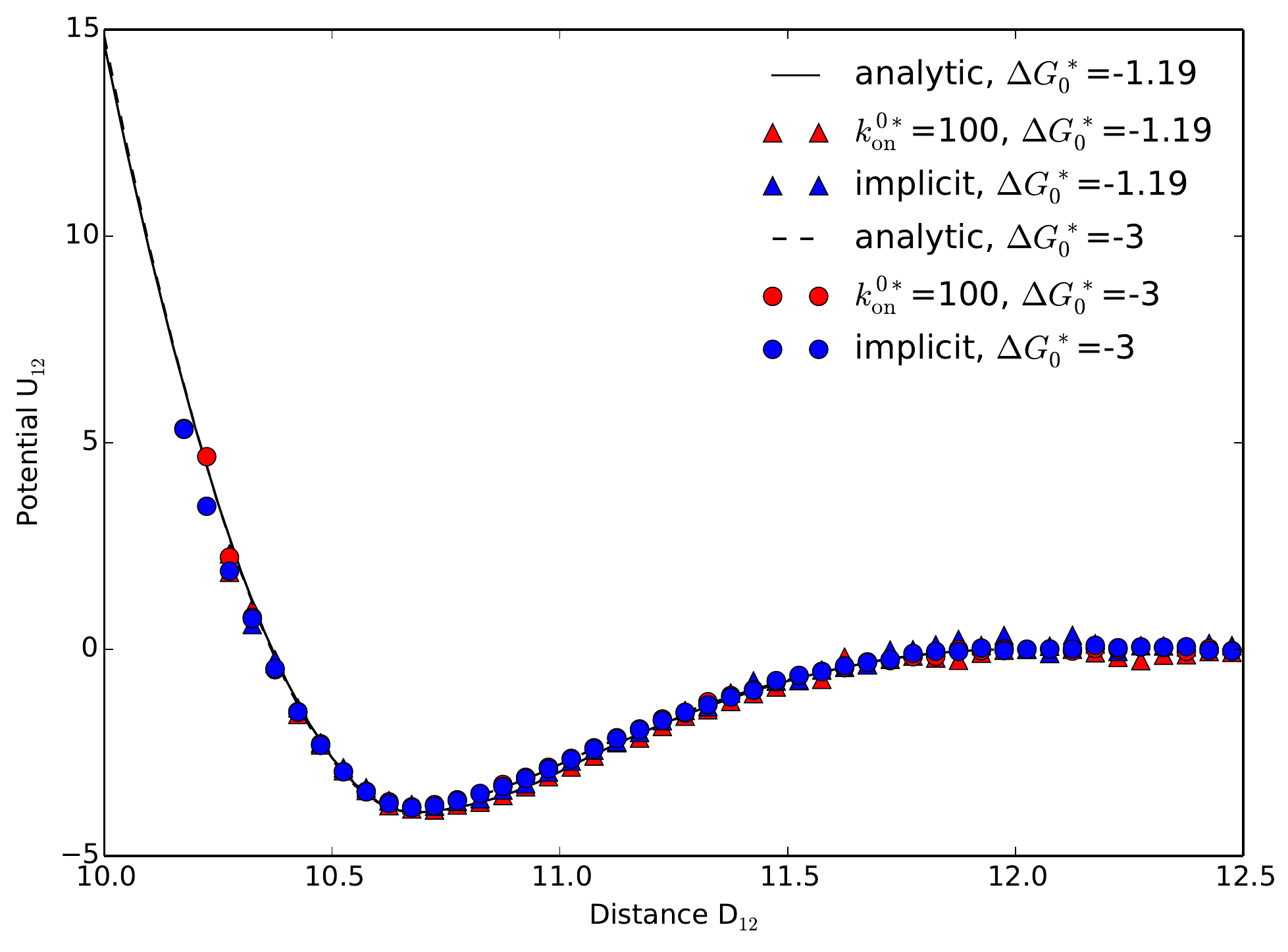} 
 \caption{Effective pair interaction as calculated using implicit-linker (blue symbols) and the explicit-linker method (red symbols) as a function of the distance between particle centres of mass ($D_{12}$). Curves refer to analytic predictions of $U_{12}$. Triangles refer to a system with $N=100$ linkers and $\Delta G_{0}^{*}=-1.19$, while dots to a system with $N=40$ and $\Delta G_{0}^{*}=-3$.  }
 \label{fig:S5}
\end{figure}

 \begin{figure}[h]
    \includegraphics[width=17cm]{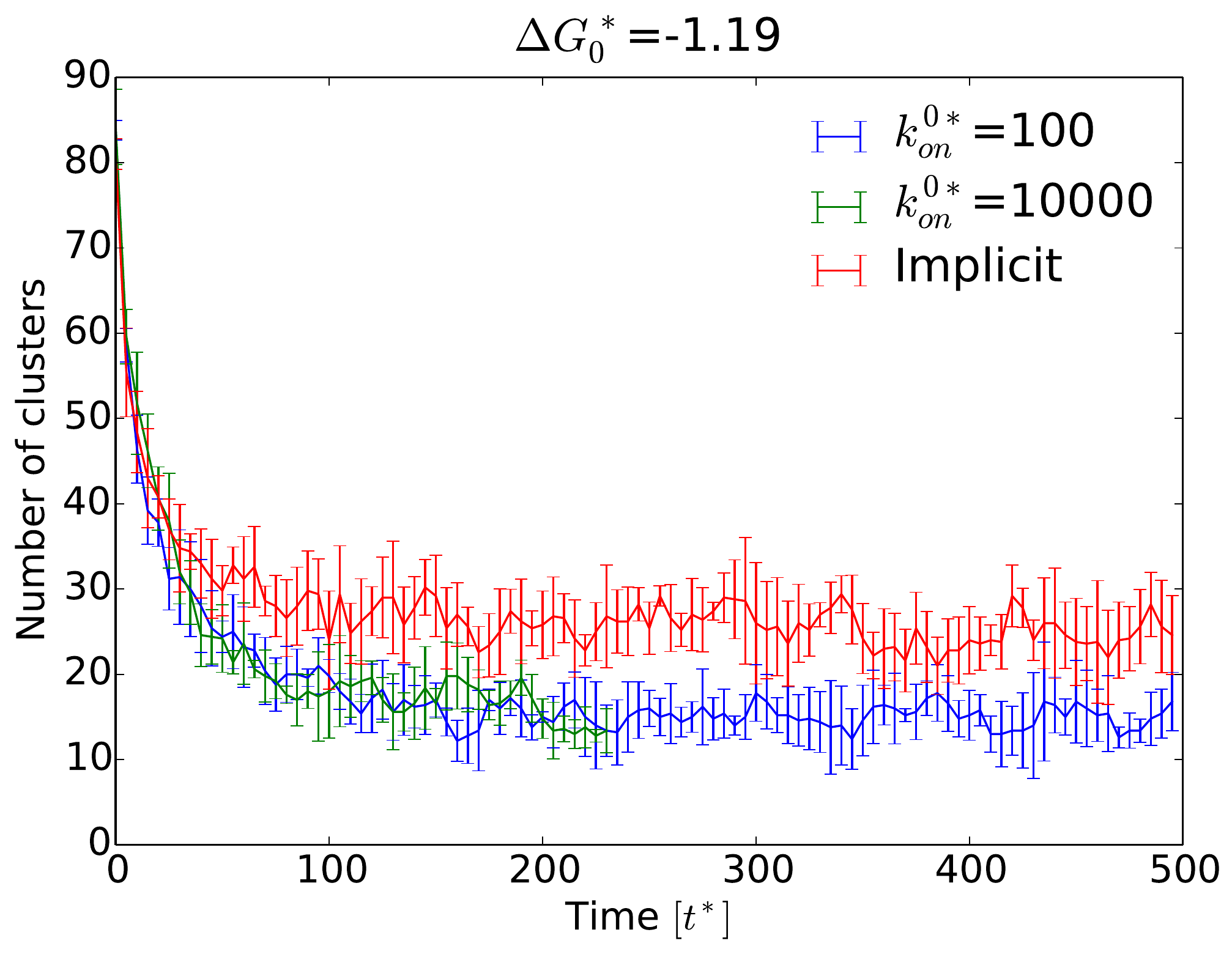} 
 \caption{Number of clusters {\em versus} simulation time for a model with $N=100$ and $\Delta G_{0}^{*}=-1.19$.
 }
 \label{fig:S4}
\end{figure}

 \begin{figure}[h]
    \includegraphics[width=17cm]{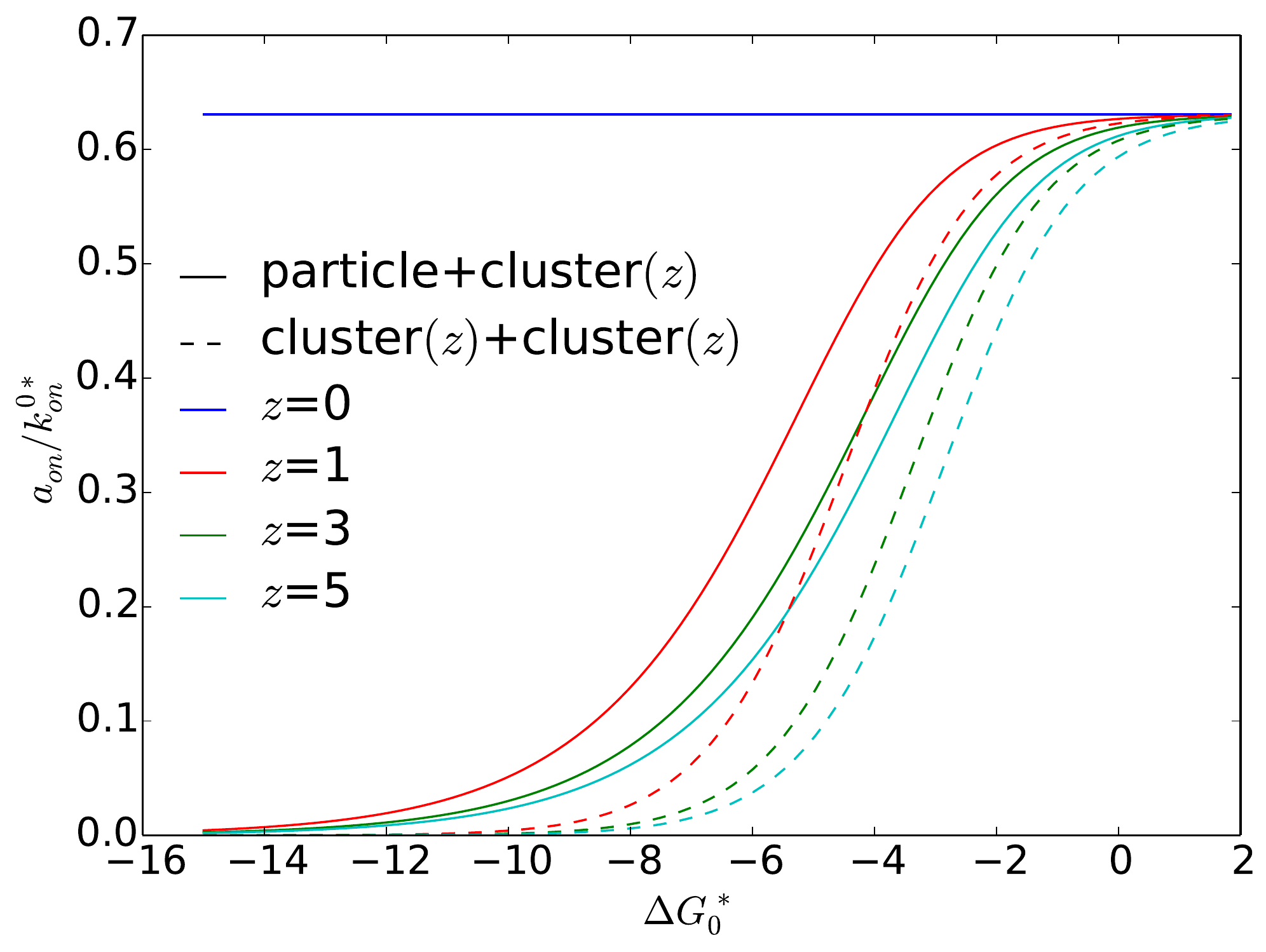} 
 \caption{Propensity (see manuscript Eq.\ 9) of attaching an isolated particle to an existing cluster (solid lines), and propensity of creating a new linkage between two particles embedded in the same cluster (dashed lines). If compared to manuscript Fig.\ 6 the number of linkers was increased from $N=40$ to $N=100$. In all calculations the distance between linkable particles has been taken equal to $2\cdot R+L$ and $z$ is the cluster coordination number.}
 \label{fig:S7}
\end{figure}

\end{document}